\documentclass[10pt, twocolumn, journal]{IEEEtran}
\IEEEoverridecommandlockouts

\usepackage{subfigure}
\usepackage{paralist}
\usepackage{color}
\usepackage{url}
\usepackage[linesnumbered,ruled,boxed,vlined]{algorithm2e}

\usepackage{algpseudocode}

\usepackage{amsmath}
\usepackage{amssymb}
\usepackage{graphicx}
\usepackage{bm}
\usepackage{paralist}
\usepackage{subfigure}
\usepackage{times}
\usepackage{balance}
\usepackage{cite}
\usepackage{comment}

\newcommand{\separator}{
  \begin{center}
    \rule{\columnwidth}{0.3mm}
  \end{center}
}

\def\ie{{\it i.e.}}

\newtheorem{definition}{Definition}

\newcommand{\beq}{\begin{eqnarray*}}
\newcommand{\eeq}{\end{eqnarray*}}
\newcommand{\beqn}{\begin{eqnarray}}
\newcommand{\eeqn}{\end{eqnarray}}
\newcommand{\bemn}{\begin{multiline}}
\newcommand{\eemn}{\end{multiline}}

\renewcommand{\ie
}{{\em i.e., }}

\newcommand\etal{{\em et al.}}

\allowdisplaybreaks
\begin{document}

\title{Detection of Rumors and Their Sources in Social Networks: A Comprehensive Survey  \\

     }

 \author
 {Otabek Sattarov$^1$ and Jaeyoung Choi$^{1*}$
\thanks{$^1$School of Computing, Gachon University, 1342 Seongnamdaero, Sujeong-gu, Seongnam-si, Gyeongi-do 13120.
 This work was supported by the National Research Foundation of Korea (NRF) grant
funded by the Korean government (MSIT) (No. 2022R1C1C1004590). $^*$ Corresponding author (email: jychoi19@gachon.ac.kr).  }
 }

\maketitle

\begin{abstract} With the recent advancements in social network platform technology, an overwhelming amount of information is spreading rapidly. In this situation, it can become increasingly difficult to discern what information is false or true. If false information proliferates significantly, it can lead to undesirable outcomes. Hence, when we receive some information, we can pose the following two questions: $(i)$ Is the information true? $(ii)$ If not, who initially spread that information?
The first problem is the rumor detection issue, while the second is the rumor source detection problem. A rumor-detection problem involves identifying and mitigating false or misleading information spread via various communication channels, particularly online platforms and social media. Rumors can range from harmless ones to deliberately misleading content aimed at deceiving or manipulating audiences. Detecting misinformation is crucial for maintaining the integrity of information ecosystems and preventing harmful effects such as the spread of false beliefs, polarization, and even societal harm. Therefore, it is very important to quickly distinguish such misinformation while simultaneously finding its source to block it from spreading on the network. However, most of the existing surveys have analyzed these two issues separately. In this work, we first survey the existing research on the rumor-detection and rumor source detection problems with joint detection approaches, simultaneously. 
 This survey deals with these two issues together so that their relationship can be observed and it provides how the two problems are similar and different. The limitations arising from the rumor detection, rumor source detection, and their combination problems are also explained, and some challenges to be addressed in future works are presented.
\end{abstract}


 \vspace{-0.5cm}
\section{Introduction}
\label{sec:intro}

A rumor is unverified or unconfirmed information circulated among people, typically through word of mouth or social-media platforms \cite{Michela2016,Skinnell2021,Juan2018}. Rumors often spread rapidly and can cover various topics, including gossip about individuals, news events, or conspiracy theories. They may contain elements of truth but are more often exaggerated, distorted, or entirely fabricated. The lack of credible evidence or confirmation characterizes rumors. They are based on hearsay, speculation, or incomplete information, making their accuracy uncertain.
Rumors typically spread quickly and informally within social networks, often fueled by curiosity, fear, or uncertainty. With the advent of social media, rumors now propagate rapidly among large audiences within a short period.
As rumors circulate, they may undergo changes, adaptations, or embellishments. Individuals may add their interpretations or details to the original rumor, leading to variations in its content.
Rumors often emerge in situations of ambiguity or anxiety, where people seek explanations or reassurance. They can reflect underlying social tensions, fears, or desires within a community.

The severity of a rumor depends on various factors, including its content, context, and impact on individuals or society \cite{Ding2022}. 
Rumors can have serious consequences, particularly if they lead to panic, social unrest, or harm individuals' reputations or well-being. For example, false health-related rumors can discourage people from seeking medical treatment or vaccination, leading to public-health risks.
Rumors originating from credible sources or authority figures may be perceived as trustworthy and can have a greater impact on public opinion or behavior.
The extent to which a rumor spreads and reaches a wide audience can amplify its impact and severity. Social-media platforms and online communication channels can facilitate the rapid dissemination of rumors to millions of people worldwide \cite{Bo2023,SOROUSH2018}.
Some rumors persist over time, becoming deeply entrenched in collective beliefs or cultural narratives. Persistent rumors can be challenging to debunk or correct, leading to long-term consequences for individuals or society.
Addressing the severity of rumors requires a multifaceted approach that involves promoting critical-thinking skills, fact-checking mechanisms, and transparent communication channels. By empowering individuals to evaluate information critically and providing accurate, timely, and transparent information, it is possible to mitigate the impact of rumors and promote information integrity within communities. In a recent situation where numerous false pieces of information like rumors spread on social networks, we can pose the following two questions: $(i)$ Is the information true? $(ii)$ If not, who initially spread that information?

\begin{figure*}[t!]
\begin{center} \centering
\includegraphics[width=0.7\linewidth]{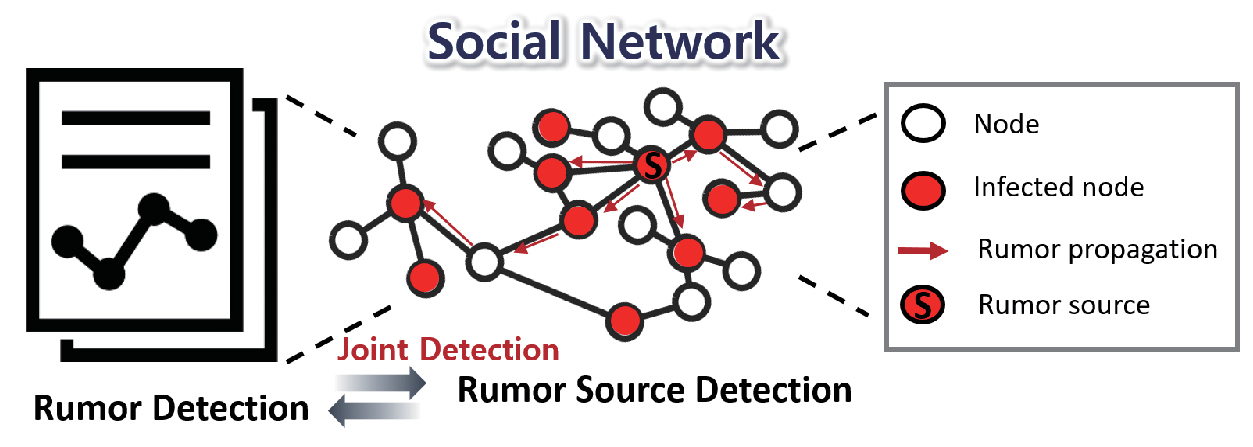}
\caption{Three main problems considered in this survey---rumor detection, rumor-source detection, and joint detection of rumor and rumor source in social networks. Here, ``Infected node" is the node that heard the rumor from its neighbors. }
\label{fig:overall}
\end{center}
 \vspace{-0.5cm}
\end{figure*}

Many studies have been conducted on rumor detection as an answer to the first question. Recent advances in social-networking platforms have led to active research on detecting rumors spreading within these networks. It is very important to detect rumors quickly to prevent the spread of false information. Here, rumor detection refers to the process of identifying and assessing the veracity of unverified or unconfirmed information circulating within a community or across communication channels \cite{Samah2018}. Rumor detection aims to distinguish between factual information and rumors, which may be inaccurate, misleading, or entirely false. Rumor detection involves analyzing various sources of information, such as social-media posts, news articles, and user-generated content, to determine the credibility and reliability of the information being shared \cite{Li2019}. This process may involve examining linguistic cues, metadata, propagation patterns, cross-referencing with trusted sources, or fact-checking databases. By detecting and verifying rumors, individuals, organizations, and platforms can take appropriate actions to mitigate the spread of misinformation and promote the dissemination of accurate information.

As an answer to the second question, much research on rumor source detection has been conducted to date. Rumor-source detection refers to the process of identifying the origin or initial source of a rumor within a network or community \cite{Jiang2016}. This can be another way to prevent rumors from spreading by identifing and controlling their sources. The primary goal of rumor-source detection is to retrace the propagation path of a rumor to its point of origin or to identify the individuals or entities responsible for its creation and dissemination. This process involves analyzing various factors, such as propagation patterns, metadata associated with shared content (e.g., timestamps, geolocation data, etc.), and network structures (e.g., social connections, information flow, etc.), to infer the source of the rumor. Rumor-source detection techniques may include propagation tracing, source attribution analysis, and metadata analysis \cite{Jiang2018,Li2023}. By identifying the source of a rumor, individuals, organizations, and platforms can better understand the dynamics of misinformation spread and implement targeted interventions to mitigate its impact and prevent future occurrences.
Rumor-source detection is clearly of practical importance because harmful diffusion can be mitigated or even blocked, e.g., 

Rumor detection and rumor-source detection are two interconnected tasks in the realm of misinformation research. While rumor detection focuses on identifying whether a piece of information is a rumor or not, rumor-
source detection aims to uncover the origin or source of the rumor. 
Jointly addressing both tasks involves integrating information from rumor and source detection to obtain a more comprehensive understanding of the rumor's characteristics and origin. 
When addressed jointly, rumor and source detection can complement each other in several ways.
Integrating information from both tasks can improve the accuracy of rumor detection by considering the content characteristics of messages and the context provided by the identified rumor sources.
Jointly analyzing rumor detection and rumor-source detection results can provide insights into the characteristics and motivations of individuals or groups behind the spread of the rumor, leading to a deeper understanding of the underlying dynamics of misinformation propagation.
By identifying both the rumors and their sources, it becomes possible to develop more effective strategies for mitigating the impact of misinformation, such as targeted interventions, corrective measures, and community-based interventions.

In this survey, we summarize the existing research on rumor detection, rumor-source detection, and joint treatments for these two aspects, as shown in Fig~\ref{fig:overall}. This distinguishes our survey from others and is expected to benefit researchers who investigate them together.

 \begin{table*}[t!]
\renewcommand{\arraystretch}{1.3}\label{tab:problem} 
\caption{Descriptions of the problems considered in this paper.}  \centering \begin{tabular}{c||c|p{3cm}|p{7cm}}
\hline      & \textbf{Problem Type} & ~~~~~~~~~~~~~\textbf{Data} &~~~~~~~~~~~~~~~~~~~~~~~~~~~~\textbf{Description} \\
\hline \textbf{Rumor Detection} & Classification & Kaggle, Twitter, Weibo, BBC news, Facebook, Web data, etc. & For a given task, the problem is determining whether the information is real or fake. \\
\hline  \textbf{Rumor-Source Detection}  & Inference & Network structure, Propagation model,  Diffusion snapshot, etc.& For a given diffusion snapshot observation, the problem is finding the node that first spreads the information in the network.\\
 \hline
 \textbf{Joint Detection}  & Joint Inference & Data on both rumor detection and rumor-source detection &For a given task and diffusion snapshot, the problem is to find out whether the information being spread over the network is a rumor or not while determining the node that first spread the information.\\
  \hline
 \end{tabular} \end{table*}

Our main contributions are as follows: 
\smallskip
\begin{enumerate}[(i)]

\item First, to the best of our knowledge, this is the first literature survey paper that deals with rumor detection, rumor-source detection, and their joint consideration. For this, we have organized the consideration of two independent problems and the problems that consider both simultaneously.

\item Second, we define the problems of rumor detection and rumor-source detection in social networks formally and organize them in an easy-to-understand manner based on related studies. Additionally, we formally define the joint detection problem in which both are considered simultaneously.

\item Third, we summarize the various algorithms for solving each problem. In the case of rumor detection, the proposed algorithms are organized into content-based, propagation-based, source-based, and hybrid approaches. Rumor-source detection is classified into single-source and multiple-source detection algorithms according to the given snapshot environment. Additionally, we have summarized related papers that consider both these approaches simultaneously.

\item Finally, we also introduce the problem of hiding rumors and rumor sources and summarize the related papers. The limitations arising from the rumor-detection, rumor-source--detection, and their combination problems are also explained, and some challenges to be addressed in future works are presented.

\end{enumerate}

The remainder of this paper is organized as follows. In Section \ref{sec:formuation}, we discuss the problem formulation for each problem. In Section
\ref{sec:algorithm}, we describe the existing algorithms for the three problems.
In Section \ref{sec:hiding}, we explain the opposite problem of how to hide the rumors and their sources in the social network, which is also a related aspect of our studies. Finally, we conclude the paper with some discussion in Section \ref{sec:conclusion}.

 \vspace{-0.5cm}
\section{Problem Formulation}
\label{sec:formuation}
In this section, we will introduce the problem formulations for the following three problems: $(i)$ rumor-detection problem, $(ii)$ rumor-source-detection problem, and $(iii)$ joint detection problem, which are briefly summarized in Table~\ref{tab:problem}. Before providing explicit descriptions, we will first explain the steps each problem comprises.

\vspace{-0.4cm}
\subsection{Rumor-Detection Problem}
 
Before explaining the rumor-detection problem in detail, we first define the term ``rumor’’, as follows.

\smallskip
\begin{definition} (\noindent{\bf Rumor})
A rumor is a piece of information or a story that is circulating widely but is unverified or of uncertain origin.
\end{definition}

In this survey, we consider that rumors can take various forms, including false news stories, misleading claims, and unverified information. In the detection step, we must analyze textual content, user interactions, and propagation patterns to identify suspicious or potentially false information. Hence, the rumor-detection problem in social networks involves identifying and mitigating the spread of false or unverified information within online communities. We organize this problem into the following three steps.


\smallskip
\subsubsection{Rumor generation}
Rumors often arise in situations of uncertainty or ambiguity, where individuals lack accurate information or understanding about a particular event, topic, or situation \cite{Michela2016,Skinnell2021}. People may create and share rumors as a way to make sense of uncertain or ambiguous circumstances and fill the gaps in their knowledge \cite{Juan2018}.
Fear, anxiety, and emotional arousal can drive the generation and spread of rumors. In times of crises, such as natural disasters, public-health emergencies, or political unrest, people may feel anxious or fearful about their safety and well-being. Rumors may emerge as a way for individuals to express and cope with their fears, seek reassurance, or warn others about potential threats.
People are naturally curious and seek information about the world around them. In the absence of reliable sources of information, individuals may turn to rumors as a way to satisfy their curiosity and gain insights into unfolding events. Additionally, some individuals may spread rumors to gain attention or social status within their social networks.
Social networks play a significant role in the spread of rumors, as individuals are influenced by the beliefs, attitudes, and behaviors of others within their social circles \cite{Samah2018}. People may spread rumors to conform to group norms, fit in with their peers, or gain approval from others. Social-media platforms amplify the effects of social influence by enabling the rapid dissemination of information to large audiences.
Further, rumors may arise in environments characterized by mistrust, skepticism, or perceived injustices. When individuals distrust authorities, institutions, or mainstream media sources, they may turn to alternative sources of information, including rumors, conspiracy theories, and misinformation. Rumors can serve as a means of expressing dissent, questioning official narratives, or challenging existing power structures.
Rumors can also be deliberately created and disseminated by individuals, organizations, or governments for strategic purposes. This may include spreading disinformation to sow confusion, discredit rivals, manipulate public opinion, or achieve political or economic objectives. Such rumors may be carefully crafted to exploit existing social divisions, exploit emotional vulnerabilities, or advance specific agendas.

Overall, the generation of rumors in social networks is a complex phenomenon influenced by a combination of individual, social, and environmental factors \cite{Ahsan2019}. Understanding these factors is essential for effectively addressing rumors and promoting information integrity within online communities.

\smallskip
\subsubsection{Rumor propagation}
Rumors spread in social networks through a combination of social interactions, technological features, and psychological factors \cite{Rachid2020}.
Rumors spread rapidly through social networks via interactions such as shares, likes, and comments. Propagation analysis involves tracking the flow of information through the network and identifying patterns indicative of rumor propagation. Key factors to be considered include the size and velocity of information cascades and user engagement levels in spreading the rumor.
Once a rumor is initiated, it spreads through social networks as users share it with their connections. Social-media platforms, messaging apps, and online forums are convenient channels for sharing information with large audiences \cite{Damian2023}.

As the rumor spreads, it may undergo amplification, with each user who shares it is contributing to its dissemination \cite{KWON2013, Qian2015}. Social-media algorithms may also amplify the spread of rumors by promoting content that elicits strong emotional reactions or generates high levels of engagement.
Rumors often gain momentum through repetition, as users encounter the same information from multiple sources within their social networks. Repetition reinforces the perceived validity and importance of the rumor, leading to further sharing and propagation.
The structure of the social network itself plays a significant role in the spread of rumors \cite{Jie2007}. Rumors are more likely to spread rapidly within densely connected communities or clusters of users with strong social ties\cite{Rajeh2021,Cherifi2019}. Information cascades can form as users share the rumor with their immediate connections, who then share it with their connections, and so on. Therefore, it is very important to understand the characteristics of the community structure of these networks.

\smallskip
\subsubsection{Rumor detection}

Rumor detection analyzes the textual content of social-media posts, news articles, and other online sources to identify linguistic cues, inconsistencies, or similarities with known rumors \cite{Damian2023, Yuan2020}. Natural language processing (NLP) techniques, sentiment analysis, and topic modeling are commonly used to analyze and classify the content of rumors.
Rumor-detection systems are often deployed as part of real-time monitoring platforms that continuously analyze social-network data for the presence of rumors. These systems can automatically generate alerts or notifications when potential rumors are detected, enabling timely intervention and response \cite{Pan2018}.
 Once a rumor is detected, various mitigation strategies can be employed to counteract its spread and minimize its impact. These strategies involve fact-checking, providing additional context or corrections, promoting media literacy and critical-thinking skills, and leveraging social-network algorithms to reduce the visibility of false information \cite{XINYI2020}.

In summary, the rumor-detection problem in social networks is a complex and dynamic challenge that requires a multidisciplinary approach, combining techniques from computer science, social-network analysis, and communication studies. By developing effective detection algorithms and mitigation strategies, it is possible to combat the spread of misinformation and promote information integrity within online communities. We formulate the rumor-detection problem (RD problem) as follows.

  \par\noindent\hrulefill
\begin{definition} (RD Problem) 
For a given task $T$, the RD problem is to find a proper classification function $f$ such that $f:T \rightarrow \{0,1 \}$, where ``$0$" indicates the rumor and ``$1$" indicates true information.
\end{definition}
  \par\noindent\hrulefill

   \vspace{-0.3cm}
\subsection{Rumor-Source--Detection Problem}
We first define a rumor source before summarizing the source-detection problem in what follows.

\smallskip
\begin{definition} (\noindent{\bf Rumor Source})
A rumor source refers to the origin or initial person, group, or entity that spreads a rumor. 
\end{definition}

Hence, the rumor source can be an individual, a media outlet, a social-media post, or any other source disseminating information.
The rumor-source-detection problem in social networks involves identifying the original source or sources of rumors within online social networks.  
This problem is an inference problem that involves finding the source of a rumor after it has sufficiently spread in the network, and much research has been done so far on this. The problem can be analyzed differently depending on the network structure and rumor propagation model being considered, and estimates are made using an appropriate estimator based on them.

\begin{figure*}[t!]
\begin{center} \centering
\includegraphics[width=0.8\linewidth]{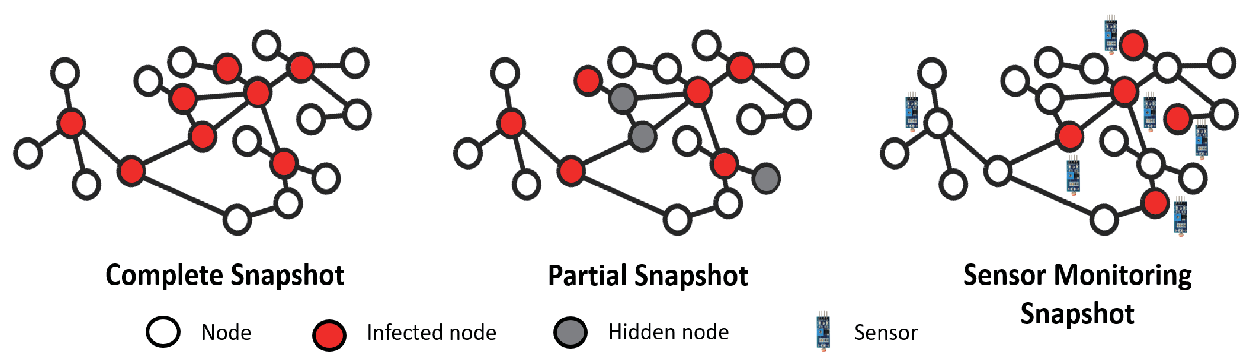}
\caption{Example of propagation snapshot. A complete snapshot means that the entire spread of the rumor can be observed. A partial snapshot means that the infection status is partially hidden when a rumor spreads. Sensor-monitoring snapshot refers to installing sensors in some nodes in the network to observe the spread of rumors.}
\label{fig:snapshot}
\end{center}
 \vspace{-0.5cm}
\end{figure*}

\smallskip
\subsubsection{Underlying network structure}

Social-network analysis techniques are used to analyze the structure and dynamics of the social network to identify potential sources of rumors. Key factors to be considered include the connectivity of users, their influence or centrality within the network, and their patterns of interactions with other users. The way rumors spread varies depending on the network structure \cite{Ganesh05}. Therefore, it is essential to have information about the type of network to identify the origin of the rumor effectively. However, in early research, because of the combinatorial complexity of inference in many cases, the network structure was often assumed to be tree-shaped \cite{shah2010,zhu2013}. This was because a tree structure has only one path connecting any two nodes, making it very effective for estimating the origin through such paths. However, in reality, the structure of social networks is generally a graph with loops, so there are limitations to approximating it as a tree.
As a result, another considered network structure is the random-graph structure \cite{shah2012}. Random graphs grow over time, and in some cases, they often have a tree-like structure from a local perspective. This makes it very convenient to analyze the path of rumors.
The final network structure under consideration is the general graph, which refers to a typical graph with a very complex structure, such as Facebook or Twitter. Here, numerous paths connect any two nodes randomly, making it very difficult to analyze how rumors spread along different paths. Therefore, most analyses approximate it using a local tree-like structure, such as the breadth-first search (BFS) algorithm \cite{shah2010tit}.

\smallskip
\subsubsection{Propagation model}

Propagation analysis involves tracking the spread of rumors through the social network to identify the patterns and dynamics of information dissemination. Understanding how rumors propagate can provide insights into potential sources and influential users who contribute to their spread.
The phenomenon of information spreading on social networks is often interpreted using epidemic models, primarily the epidemic model for the spread of diseases in arbitrary contact networks \cite{Li2017}.
This is because epidemic disease in the population is similar to rumor diffusion in a social network. 
In the susceptible--infected (SI) model, a node initially exists in a susceptible (S) state, meaning it can be infected (I). When the node contacts with an infected neighbor, it becomes infected, transitioning from the susceptible state to the infected state.
 From the perspective of rumor propagation in social networks, an infected node is the one that has received the rumor. A susceptible node is a node that has not received any rumor; however, because of its neighboring infected nodes, it can become infected after receiving a rumor \cite{Sushila2018}. The SI model assumes that once a node is infected, it remains infected indefinitely. This limitation makes it less suitable for modeling real-world scenarios such as contagious diseases where individuals can recover over time or with proper treatment.
Models including the susceptible--infected--recovered (SIR) model and susceptible--infected--susceptible (SIS) model have been proposed to address such issues. 
In the SIR model, susceptible nodes become infected over time but eventually recover with either the passage of time or acquired immunity. Furthermore, the SIR model assumes that once a node has recovered, it will never be infected again. However, such an assumption is not always appropriate in reality, as immunity can sometimes wane over time. 
A model that addresses this issue is the susceptible—infected—recovered--susceptible (SIRS) model. This model is more realistic as it allows individuals to return to a susceptible state after recovery, reflecting the possibility of waning immunity observed in reality.

As an information-spreading model, Kempe \etal \cite{Kempe03} first considered a well-known {independent cascade} (IC) model, whose the description is as follows.
In this model, nodes have three possible states---susceptible (S), active (A), and inactive (I). In the susceptible state, a node
is allowed to be activated.
An active state is
one that is activated in the previous time slot and can
activate other susceptible child nodes. An inactive state
denotes a state activated earlier but cannot
activate other susceptible nodes anymore. The activated nodes are
active for only one-time slot and become inactive at the next
time slot. Once a node becomes inactive, it maintains the state until
the end of the cascade process. More precisely,
if a node $i$ receives one information at
time $t$ from one of its infected nodes, it tries to spread its own information to its neighbor $j$ with
probability $q_{ij}$ at the next time $t+1$.
Ok \etal \cite{Ok2014} explained the phenomenon of information spreading throughout the network using a game diffusion model. They demonstrated how information spreads by modeling the reward individuals receive for adopting a strategy that dictates whether they accept information spreading from neighboring nodes, considering whether they believe it or not.

\smallskip
\subsubsection{Snapshot observation}
After a rumor spreads in a network, the problem of identifying its origin ultimately depends on when and how the administrator observed snapshots of the information spreading in the network \cite{shah2012,zhu2013}. For instance, if the administrator observed the network at the onset of the rumor spreading, it would likely be much easier to trace its origin compared to observing it after it spread extensively. Alternatively, it can be anticipated that identifying the origin becomes easier when all the disseminated information is known than when only a portion of the disseminated information is given. Therefore, methods of observing snapshots in a network are generally categorized into three types, as depicted in Fig~\ref{fig:snapshot}.

\smallskip
\begin{enumerate}[(i)]

\item Complete snapshot: The complete-snapshot observation considered in this paper refers to an observation where all infected nodes can be identified without being concealed or disappearing because of any other reason while the rumor spreads in the network \cite{shah2010tit}, as depicted in the leftmost figure in Fig~\ref{fig:snapshot}. It refers to a scenario where there are no instances of concealing one's infection status, except for cases where recovery from infection occurs and cannot be distinguished.

\item Partial snapshot: In contrast to the complete snapshot, this scenario refers to situations in the network where infected nodes cannot be identified owing to various reasons \cite{Leng2014}. These could involve infected nodes deliberately hiding their infection status or the introduction of noise in infection situations where infected nodes probabilistically report their status differently, as shown in the middle figure in Fig~\ref{fig:snapshot}. In such cases, as the true status of hidden infected nodes needs to be estimated to analyze the infection pathways, the typical approach tends to be more complex.

\item Sensor-monitoring snapshot: In this case, rather than the administrator observing the phenomenon of rumors spreading throughout the network, the approach involves receiving information about the spread of rumors from reports of a few sensor nodes planted in the network, as depicted in the rightmost figure in Fig~\ref{fig:snapshot}. The process typically involves initially planting several sensor nodes in the network \cite{Brunella2019}. As information spreads, these nodes collect data on who transmitted what information and when and then relay this information to the administrator.
Adding more sensor monitor can improve the detection accuracy while it may decrease the performance of the system
\cite{Sushila2018}.

\end{enumerate}

The algorithm for identifying rumor sources can vary depending on the type of snapshot information provided. Therefore, it is first necessary to distinguish the type of information provided.

\smallskip
\subsubsection{Estimator}
\begin{figure}[t!]
\begin{center} \centering
\includegraphics[width=1\linewidth]{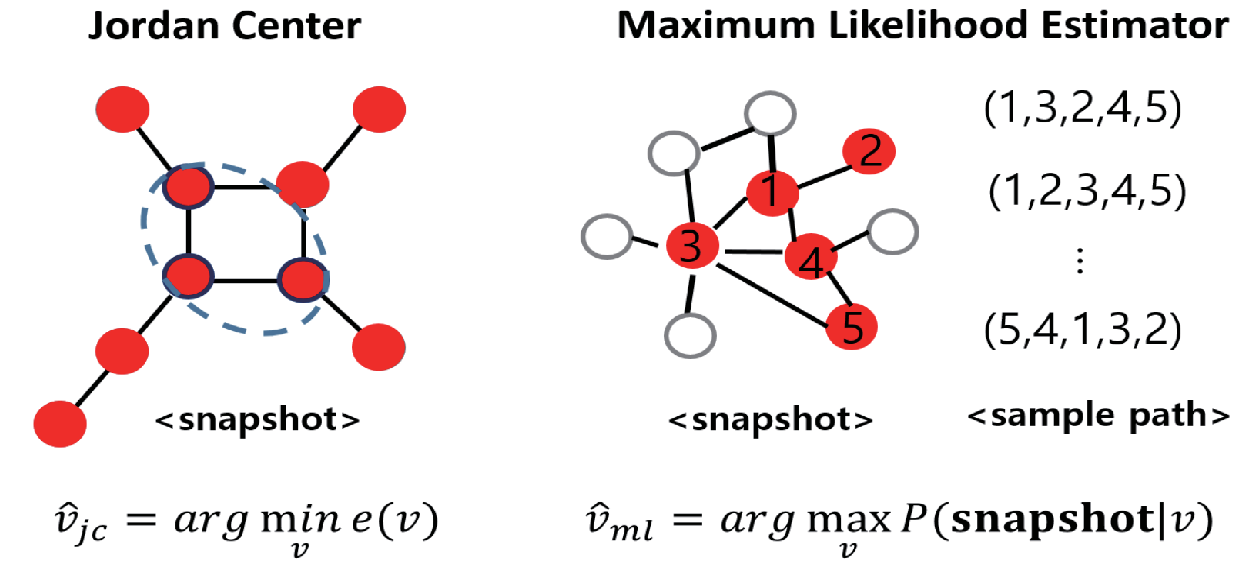}
\caption{Examples of Estimators. The estimator on the left is the Jordan center, one of the graph-based estimators. The estimator on the right is the MLE, one of the probabilistic estimators. While the graph-based estimator estimates the center node by analyzing the snapshots of all rumors, the probabilistic estimator calculates the probability for the sample path of the infection route and selects the node with the maximum.}
\label{fig:estimator}
\end{center}
 \vspace{-0.2cm}
\end{figure}

 Two kinds of approaches are considered as estimators: $(i)$ graph-based estimator and $(ii)$ probability-based estimator, as shown in Fig~\ref{fig:estimator}. 
A graph-based estimator refers to an estimator obtained by applying graph theory-based centrality based on infected or recovered nodes on the graph.
Probability-based estimator refers to an estimator obtained by calculating the probability of the infection path of an infected or recovered node.

\smallskip
\noindent{\bf Graph-Based Estimator.}
For a given propagation snapshot of the rumor spread over the network when observing a diffusion snapshot on the network at a random time, it is predicted that the central node of the infection graph will be the source.  Therefore, we can consider using centrality measures as an estimator. In a graph $G,$ and for a node $v$ in $G$, the centralities exist such as Jordan center or eigenvalue center, etc.

\smallskip
\noindent{\bf Maximum Likelihood Estimator.}
As an estimator of the source, we first consider an MLE for the observed graph (snapshot) $G_N$ when there are $N$ infected nodes in the network:
\begin{equation}\label{eqn:MLE}
\begin{aligned}
\hat{v}_{\tt ml}&=\arg \max_{v\in G_N}P(G_{N}|v),
\end{aligned}
\end{equation}
\i.e., the MLE is the node that maximizes the likelihood $P(G_{N}|v)$ of the diffusion snapshot $G_N$.
To compute the MLE, we let $\sigma$ be a sample path that generates the snapshot $G_N$. Then, for a given $G_N,$ there are a large number of 
possible sample paths, which makes the MLE computation an NP-hard problem \cite{shah2010}. However, it has been found that this problem can be analyzed theoretically with a specific centrality characteristic \footnote{In many cases, these probabilistic approaches were directed to existing graph centers such as Jordan center \cite{zhu2013} or new graph centers such as Rumor center \cite{shah2010}. } when the degree of all nodes is the same as that of a regular tree and the homogeneous diffusion rate \cite{shah2010}. Instead of calculating the probabilities of all sample paths, another method has been proposed to calculate the sample path with the highest probability and estimate its first node as the source \cite{zhu2013}.

The rumor-source-detection problem (RSD problem) can be formulated from this graph-based or probability-based estimator as follows.

  \par\noindent\hrulefill
\begin{definition} (RSD Problem) 
For a given infection graph $G_I = (V_I, E)$, where $V_I$ is the infected node set, the RSD problem is to find a proper estimation function $f$ such that $f:G_I \rightarrow \widehat{S}$, where $\widehat{S} \subseteq V$ is the estimator of the true rumor source set $S^*$.
\end{definition}
  \par\noindent\hrulefill

 \vspace{-0.5cm}
\subsection{Joint Detection Problem}
\label{}
The joint inference of rumor and rumor-source detection involves simultaneously identifying both the existence of a rumor and its origin within a network or community. This approach integrates techniques from rumor detection and source attribution to obtain a comprehensive understanding of the rumor-propagation process. From what is known so far, this joint problem can be explained by the following two reasons. 

\smallskip
\subsubsection{Source reliability}
In social networks, the credibility of information transmission is closely related to the credibility of the nodes spreading that information. For instance, if someone consistently spreads false information, it is likely that any further information they spread will also be viewed as false. Conversely, if someone who mostly shares truthful information spreads additional information, the information is likely to be considered highly credible. Therefore, when assessing the truthfulness of information spreading in the current network, it would be helpful to estimate who propagated the information, along with assessing the reliability of that source. Hence, the estimation of both the origin and truthfulness of information occurs simultaneously, leading to a joint inference problem.

\smallskip
\subsubsection{Number of sources}
It is known that \cite{Seo2012} rumors are usually initiated by a small number of people. In contrast, true information
is reported by many people independently. Therefore, when the number of nodes spreading a particular piece of information is small, it can be deemed a rumor. On the contrary, if many people independently spread information, it can be considered as truth rather than a rumor. In this case, instead of assuming a single rumor source, it is generally assumed that there could be multiple sources. Afterward, based on the data of the spread information, it is necessary to jointly estimate both the number of sources and the sources themselves and then use this information as a basis for jointly estimating the truthfulness of the information.

Therefore, the joint detection problem (JD problem) can be formulated by considering rumor detection and rumor-source detection simultaneously, as follows.
  \par\noindent\hrulefill
\begin{definition} (JD Problem) 
For a given task and infection graph $(T, G_I)$, the JD problem is to find a proper detection function $f$ such that  $f:(T, G_I) \rightarrow (\{0,1 \}, \widehat{S})$, which contains the classification (RD problem) and estimation (RSD problem), simultaneously. 
\end{definition}
  \par\noindent\hrulefill

In the next section, we will introduce various algorithms that solve each of the problems defined above.
 \vspace{-0.3cm}
\section{Detection Algorithms}
\label{sec:algorithm}

\subsection{Rumor-Detection Algorithms}
Rumor-detection algorithms are computational methods used to automatically identify and classify rumors within social media and other online communication platforms. For easy understanding, we describe a simple pseudocode in Algorithm~\ref{alg:rumor}. These algorithms leverage various techniques from machine learning, NLP, network analysis, and data mining to analyze large volumes of textual and network data and distinguish between true and false information. We first classify rumor detection algorithms and techniques in Table \ref{tab:rumord} and summarize the detailed explanation of some algorithms and techniques in Table \ref{tab:rumordss} as follows.

\smallskip
\subsubsection{Content-based approaches}
Content-based methods are classified into three approaches---
knowledge-based, style-based, and multimodal-based approach as in \cite{XINYI2020,Shen2023}.
Knowledge-based rumor detection complements data-driven approaches by providing a qualitative assessment of rumors based on domain expertise, external verification, and critical analysis. By integrating domain-specific knowledge with rigorous fact-checking methodologies, knowledge-based detection enhances the accuracy and reliability of rumor-detection efforts.
Hu \etal \cite{Hu2021} proposed a novel end-to-end graphical neural model CompareNet, which compared the news with external knowledge through entities for fake news detection. For this, they constructed a directed heterogeneous document graph
containing topics and entities for each type of news. Then, they developed heterogeneous graph attention networks to learn topic-enriched news representations. Finally, they use an entity comparison network 
for fake news classifiers.
Mayank \etal \cite{Mayank2022}  proposed DEAP-FAKED, a dual-part knowledge-graph (KG) -based framework, which consisted of the three components: $(i)$ news encoder, $(ii)$ entity encoder, and $(iii)$
classification layer. First, it performed contextual encoding of the news title and then identified the title of the news by KG. Lastly, it performed the final fake news classification. They used NLP-based technology to encode the headline of the news. Then, they adopted named entity recognition (NER) to
recognize and extract named entities from the texts, and use named entity disambiguation
to map the entities to the KG.
Pan \etal \cite{Pan2018} proposed an approach for fake-news detection based on the article content by constructing knowledge graphs using
the TransE and B-TransE methods. They used three different knowledge graphs to generate background knowledge and then adopted the B-TransE method to establish the entity
and relation embedding and check whether the news articles were authentic. They found that a binary TransE model that combined positive and negative individual models performed better than a single one.

\begin{table*}[t!]
\renewcommand{\arraystretch}{1.3}
\caption{Taxonomy of Rumor Detection Algorithms.} \label{tab:rumord} \centering \begin{tabular}{p{3cm}||p{5cm}|p{7cm}}
\hline      & ~~~~~~~~~~~~~~~~~~~~~~~~\textbf{Datasets} & ~~~~~~~~~~~~~~~~~~~~ \textbf{Detection Algorithms} \\
\hline \textbf{Content-based approaches} & Kaggle, CoAID, LUN, SLN, Kaggles + BBC news, FA-KES, ISO, PolitiFact, BuzzFeed, Buzzfeed, Random political, WELFake, FNC, Weibo dataset, FakeNewsNet, Tweet, Fakeddit & DEAP-FAKED \cite{Mayank2022}, CompareNet \cite{Hu2021}, B-TransE  \cite{Pan2018}, CNN, RNN \cite{Nasir2021}, Supervised classifiers \cite{Zhou2020}, Sequential neural network \cite{Choudhary2020}, WELFake \cite{Verma2021}, CNN-LSTM \cite{Umer2020}, Multi-domain Visual Neural Network \cite{Qi2019}, Crossmodal Attention Residual and Multichannel CNN \cite{Song2020}, LIWC, KNN\cite{Singh2021}, Multimodal-VAE \cite{Khattar2019}, Fine-grained classification \cite{Kegura2022} Expert-based fact-checking \cite{Dongsong2016}, Crowdsourcing \cite{Mitra2015}
\\

\hline  \textbf{Propagation-based approaches}  & Twitter, PolitiFact, GossipCop, FbMultiLingMisinfo, Weibo, FakeNewsNet, Twitter15, Twitter16 &  Propagation2Vec \cite{Silva2021}, Geometric DL \cite{Monti2019}, Deep Error Sampling \cite{Barnabò2023}, Dynamic Graph Neural network \cite{Song2022}, Hypergraph NN \cite{Jeong2022}, GNN \cite{Han2020}, UniPF \cite{Wei2022}, Point process model \cite{Murayama2021}, Multi-View Attention Networks \cite{Ni2021}, Hierarchical propagation network \cite{Shu2020}, Stance network \cite{Davoudi2022},  TriFN \cite{Shu2019}, FANG \cite{Nguyen2020}, SVM\cite{Carlos2011}, Decision tree \cite{KWON2013} Diffusion pattern \cite{Vosoughi2018}\\
 \hline
 \textbf{Source-based approaches}  &Buzzfeed news, PolitiFact, Twitter15, Twitter16, Weibo, RumorEval19, Pheme, Presented by others Twitter, MIB, GossipCop   & Authors features \cite{Sitaula2020}, Structure-aware Multi-head Attention Network \cite{Yuan2020}, BiGRU, AGWu-RF \cite{Luvembe2023}, Contextual-LSTM \cite{Kudugunta2018}, Minimum Redundancy Maximum Relevance \cite{Rostami2020}, CNN, bi-SN-LSTM \cite{Gao2020}, Multi-view co-attention network \cite{Bazmi2023}, FAKE-DETECTOR \cite{Zhang2018}, Source credibility \cite{Viviani2017}, Malicious user detection \cite{Varol2017}\\
  \hline
   \textbf{Hybrid and other machine learning-based approaches}  & Twitter, URL, Wiki, HTML, MSN, RSS feeds, PHEME, COVID-19, Weibo & Spam fact-checking \cite{Baly2018}, Spam filter \cite{Fetterly2005}, Webpage spam filter \cite{Ntoulas2006}, Graph-based approach\cite{Benjamin2019}, Graph-regularization \cite{Abernethy2010}, VRoC\cite{Cheng2020}, GAN\cite{Cheng2021}, RDLNP\cite{Lao2021}, COVID-19\cite{Mingxi2021}, Bi-GCN \cite{Bian2020}, Graph Neural Network \cite{Liu2024}, MultiModal \cite{Peng2023}, MultiModal \cite{Zhang2024}, Genetic \cite{Somya2020}\\
  \hline
 \end{tabular} \end{table*}

  \begin{table*}[t!]
\renewcommand{\arraystretch}{1.3}
\caption{Summary of Rumor Detection Algorithms.} \label{tab:rumordss} \centering \begin{tabular}{p{2cm}|p{2cm}|p{1.5cm}|p{5cm}|p{5cm}}
\hline   & ~~\textbf{Ref} & ~~\textbf{Methods}  & ~~\textbf{Pros} & ~~\textbf{Cons} \\
\hline \textbf{Content-based approaches} & DEAP-FAKED \cite{Mayank2022} & Knowledge-based & Effective results are obtained by appropriately using NLP and tensor decomposition model to encode news content and embed KG & Lack of different result graphs for different test data sets
\\
\cline{2-2} \cline{3-2} \cline{4-2} \cline{5-2} & CompareNet \cite{Hu2021}& Knowledge-based & Develop efficient deep learning model with KG and entity description & 
Lack of efficient method of combining multi-modal data and external knowledge \\
\cline{2-2} \cline{3-2} \cline{4-2} \cline{5-2} & B-TransE  \cite{Pan2018}& Knowledge-based & Development of an algorithm that can detect fake news even with incomplete KG & 
Lack of explanation as to why performance is good even with incomplete kGs \\
\cline{2-2} \cline{3-2} \cline{4-2} \cline{5-2} & CNN, RNN \cite{Nasir2021}& Style-based & Efficient algorithm design by appropriately combining CNN and RNN & 
Using a relatively old comparison models\\
\cline{2-2} \cline{3-2} \cline{4-2} \cline{5-2} & Sequential neural network \cite{Choudhary2020} & Style-based & Proposal of a linguistic model that efficiently uses language-driven features & 
Lack of performance comparison with other models\\
\cline{2-2} \cline{3-2} \cline{4-2} \cline{5-2} & Multimodal-VAE \cite{Khattar2019} & Multimodal-based & 
Development of an efficient detection algorithm using only tests or images from social media posts & 
Lack of diverse experiments on detection results\\
\cline{2-2} \cline{3-2} \cline{4-2} \cline{5-2} & Fine-grained classification \cite{Kegura2022} & Multimodal-based & 
Performance analysis of fine-grained detection model using unimodal and multimodal simultaneously & 
Insufficient pre-training model for visual representation generation\\

\hline  \textbf{Propagation-based approaches}  & Propagation2Vec \cite{Silva2021} & Cascade-based & Jointly consideration of informative modes and cascades with propagation patterns & Works well only in specific domains and lacks scalability\\
\cline{2-2} \cline{3-2} \cline{4-2} \cline{5-2} & Dynamic Graph Neural Network \cite{Song2022} & Cascade-based & 
Extending the fake news detection problem to dynamic networks &  Homogeneous discrete-time dynamic news propagation  \\
\cline{2-2} \cline{3-2} \cline{4-2} \cline{5-2} & Hypergraph NN \cite{Jeong2022} & Cascade-based & 
Consider a hyper-graph that can explain news propagation at the group level &  Type information of each hyperedge is not considered in the framework  \\
\cline{2-2} \cline{3-2} \cline{4-2} \cline{5-2} & Multi-View Attention Networks \cite{Ni2021} & Propagation graph-based & 
Development of multi-view attention network using text semantic attention and propagation structural attention &  Response comments from news are not taken into account \\
\cline{2-2} \cline{3-2} \cline{4-2} \cline{5-2} & Hierarchical propagation network \cite{Shu2020} & Propagation graph-based & 
Hierarchical propagation network design that can implement macro-level and microlevel of fake news and true news &  Structures of hierarchical propagation networks are not considered in user proagation learning \\
\cline{2-2} \cline{3-2} \cline{4-2} \cline{5-2} & FANG \cite{Nguyen2020} & Propagation graph-based & Scalable in training and efficient at inference time without the need to re-process the entire graph &  Errors from upstream tasks can occur \\
 \hline
 \textbf{Source-based approaches}  & Authors features \cite{Sitaula2020} & Author-based & Increased detection performance with an algorithm based on rumor source credibility & It is difficult to apply if we do not know who the source or author is\\
 \cline{2-2} \cline{3-2} \cline{4-2} \cline{5-2} & Structure-aware Multi-head Attention Network \cite{Yuan2020} & Author-based & Algorithm that simultaneously considers news content, publishing, and reposting relations of publishers and users &  Results for detection experiments do not vary \\
 \cline{2-2} \cline{3-2} \cline{4-2} \cline{5-2} & BiGRU, AGWu-RF \cite{Luvembe2023} & Author-based & Improved algorithm performance through extraction of long-range context information &  Specific emotion is not considered in the model \\
  \cline{2-2} \cline{3-2} \cline{4-2} \cline{5-2} & Contextual-LSTM \cite{Kudugunta2018} & User-based & Development of an LSTM algorithm that can consider both content and metadata at the tweet level &  Detects the nature of an account from only a single tweet \\
   \cline{2-2} \cline{3-2} \cline{4-2} \cline{5-2} & FAKE-DETECTOR \cite{Zhang2018} & User-based & Designing a new algorithm that can overcome unknown characteristics of fake news and diverse connections among news articles, creators and subjects &  The data set used in the experiment is not sufficient \\
     \cline{2-2} \cline{3-2} \cline{4-2} \cline{5-2} & Malicious user detection \cite{Varol2017} & User-based & High accuracy and and can detect bots of different nature &  Limitations of experimental proofs on Twitter data \\
  \hline
   \textbf{Hybrid and machine learning-based approaches}  & Spam fact-checking \cite{Baly2018} & Contents and source- based  & To improve the performance of rumor detection, methods such as fact checking, stance detection, and source reliability are used harmoniously & The factuality and bias of the joint model are not explained well\\
     \cline{2-2} \cline{3-2} \cline{4-2} \cline{5-2} & Spam filter \cite{Fetterly2005} & Contents and source- based & Development of technology to find instances such as slices and dice in Webpage &  Lack of results for various performance metrics \\
       \cline{2-2} \cline{3-2} \cline{4-2} \cline{5-2} & Graph-based approach\cite{Benjamin2019} & Contents and propagation-based & Algorithm considering content sharing situations in communities &  Lack of interpretation of various data generated from graphs \\
         \cline{2-2} \cline{3-2} \cline{4-2} \cline{5-2} & Graph-regularization \cite{Abernethy2010} & Contents and propagation-based & Efficient algorithm design to detect spam hosts on the web through graph regularization & Insufficient performance comparison with various graph learning methods  \\
  \hline
 \end{tabular} \end{table*}

By focusing on the linguistic and stylistic aspects of rumors, style-based detection provides an additional layer of analysis that complements other detection methods. By identifying linguistic cues, semantic inconsistencies, and stylistic anomalies, style-based detection helps uncover the subtle characteristics of misinformation and improve the effectiveness of rumor-detection efforts.
Zhou \etal \cite{Zhou2020} proposed a theory-driven
Fake-news detection method that investigated news content at various levels: lexicon-level, syntax-level, semantic-level, and discourse-level. 
They used supervised classifiers such as logistic regression , naïve Bayes, support vector machine, random forests, and XGBoost.
Nasir \etal \cite{Nasir2021}  proposed a model that combined CNN and RNN for fake-news classification. They showed that the model achieved higher detection results than other nonhybrid baseline methods on two fake news datasets (ISO and FA-KES).
Umer \etal \cite{Umer2020} proposed a model that automatically classified news articles with stance labels based on titles and news text. 
They especially used principal
component analysis and the Chi-square test to extract features for the CNN-LSTM model. 
Their experimental results revealed that the proposed model improved the F1-score and accuracy by 20 and
4\%, respectively, when used with a reduced feature set.
Choudhary \etal \cite{Choudhary2020} proposed a linguistic model to extract the properties of content that generated language-driven features such as syntactic,
grammatical, sentimental, and readability features of specific news. As this model required time-consuming features, they considered a neural-based sequential learning model for rumor detection. 
Verma \etal \cite{Verma2021} proposed a machine-learning classification named WELFake model, which was a two-phase
benchmark model based on word embedding
(WE) over linguistic features for fake-news detection. The first phase was a preprocess consisting of the
data set and validated the veracity of the news content by using
linguistic features. The second phase involved merging the linguistic feature
sets with WE and applying voting classification.
Qi \etal \cite{Qi2019} suggested a multidomain visual neural network (MVNN) framework to extract the visual information of frequency and pixel domains for detecting rumors. For this, they first designed a CNN-based network model to automatically capture the patterns of fake news images in the frequency domain and then used a multibranch CNN--RNN model to extract visual features in the pixel domain.

By leveraging information from multiple modalities, multimodal rumor detection provides a more holistic and nuanced approach to identifying and classifying rumors. By integrating textual, visual, and metadata features, multimodal models capture rich and diverse information that enhances the effectiveness of rumor-detection efforts.
Khattar \etal \cite{Khattar2019} proposed an end-to-end network, multimodal variational autoencoder (MVAE), which combined a bimodal variational autoencoder with a binary classifier for rumor detection. A variant autoencoder was used for multimodal representation, and the input was
the text containing the post and attached image.
Singh \etal \cite{Singh2021} proposed a multimodal approach for detecting fake news with text
by identifying multiple text and visual features associated with fake and credible news articles.
Segura--Bedmar \etal \cite{Kegura2022} combined unimodal and
multimodal methods to detect rumors by performing a fine-grained classification. They showed that
exploiting both text and image data significantly improved the performance of fake-news detection.
Song \etal \cite{Song2020} designed a multimodal fake-news--detection framework based on crossmodal attention residual and multichannel convolutional neural networks (CARMN).  It was based on two models:  $(i)$ crossmodal attention residual network (CARN) and $(ii)$ multichannel convolutional neural network (MCN). The former was used to fuse the relevant information between different modalities and keep the unique properties for each modality, and the latter mitigated the influence of noise information simultaneously.

\smallskip
\subsubsection{Propagation-based approaches}

Rumors often spread through information cascades, where one user shares the rumor with others who, in turn, propagate it further. Algorithms can analyze the propagation patterns of information cascades on social networks to identify rumors and estimate their credibility based on factors such as the cascade size, velocity, and depth. Rumor-detection algorithms analyze the structure and dynamics of social networks to identify patterns indicative of rumor propagation. By examining the flow of information through the network, these algorithms can identify influential nodes or communities that are likely to spread rumors. 
Many studies have revealed that rumors and truth spread differently on social networks \cite{KWON2013,Monti2019}. 
The performance of graph neural networks (GNNs) has recently improved, and GNNs are now widely used in rumor-detection problems.
Silva \etal \cite{Silva2021}
proposed Propagation2Vec, which was an early detection model for fake news. 
The model first assigned varying levels of importance to the nodes and cascades in propagation networks. Then, it reconstructed the knowledge of complete propagation graphs based on their partial propagation networks at an early detection threshold. For this, they adopted a hierarchical attention mechanism to
encode the propagation network, which could assign corresponding weights to different
cascades. 
Monti \etal \cite{Monti2019} designed a novel automatic rumor-detection framework based on geometric deep learning. 
It was a generalization of CNNs to graph models considering heterogeneous data such as content, user profile and activity, social graph, and news propagation. 
Barnabò \etal \cite{Barnabò2023} considered active learning (AL) strategies for GNN for rumor detection. Then, they proposed deep error sampling, which was a new deep AL architecture. 
The key operation was that when coupled with uncertainty sampling, it performed equally well or better than the most common AL strategies and was the only existing AL procedure specifically targeting rumor detection. 
Jeong
\etal \cite{Jeong2022} designed a leverage hypergraph to represent the group-wise interactions among news by focusing on important news relations with its dual-level attention mechanism. For this they compared a propagation tree and a hypergraph and then constructed a hypergraph for the news relations by considering the user, time, and entity.
Song \etal \cite{Song2022} proposed a rumor-detection method based on a dynamic propagation graph to capture the missing dynamic propagation information in static networks. They proposed a dynamic GNN for fake news detection (DGNF) for a fake-news detection architecture based on a dynamic propagation network. 
Han \etal \cite{Han2020} considered the difference patterns between rumor and true information and adopted GNNs to deal with non-Euclidean data. To train GNNs, they used continual learning models, which showed that the proposed model presented robust performance across different datasets.
Wei \etal \cite{Wei2022} studied a new concept involving a propagation forest to cluster a tree of semantic and structural characteristics. Based on this concept, they designed a novel unified propagation forest-based framework (UniPF) to improve rumor detection for the latent correlations between propagation trees. 
Murayama \etal \cite{Murayama2021} proposed a point process model for rumor propagation on
Twitter. In the model, the dissemination of fake news consisted of two-stage processes such as the
cascade of original news and the cascade of assertion of news falsehoods. The experiments
indicated that the proposed method contributed to understanding the dynamics of the
spread of rumors on social media.

Propagation-graph-based rumor detection is a powerful framework for understanding the dynamics of information propagation and identifying rumors within social networks. By analyzing the structure, dynamics, and anomalies of propagation graphs, these approaches offer insights into the mechanisms of rumor diffusion and enable effective detection and mitigation strategies.
Ni \etal \cite{Ni2021} developed a new NN model, named multiview attention networks (MVAN), for detecting the rumor.
The MVAN model was based on source tweets and their propagation structures. For this, they adopted 
graph attention networks (GATs) to encode and represent the
propagation structure of news. The model was highly robust for early rumor detection
and exhibited some interpretability.
Shu
\etal \cite{Shu2020} proposed a hierarchical propagation network feature for rumor detection. 
They also performed a comparative statistical analysis to distinguish rumors from true information. 
They showed that extracting hierarchical features was helpful and could outperform the existing methods.
Davoudi \etal \cite{Davoudi2022} developed an automated system dynamic analysis, static analysis, and structural analysis (DSS) for the early detection of rumors wherein the propagation tree and stance network were leveraged simultaneously and dynamically.
For constructing the stance network, a novel method was proposed, and 
various graph-based feature extraction methods were utilized for sentiment analysis.
Shu \etal \cite{Shu2019} suggested that social context was more important than the content of the rumor. For this, they proposed
a tri-relationship-embedding framework, which modeled both publishers-news
relationships and user-news interactions for rumor detection. 
Nguyen
\etal \cite{Nguyen2020} designed a fact news graph (FANG), a novel graph-representation learning 
model. It was  scalable in training as it did not have to maintain all nodes and
FANG exhibited significantly better rumor detection than other recent graphical and nongraphical models.

\smallskip
\subsubsection{Source-based approaches}

Rumor-detection algorithms assess the credibility of news sources based on factors such as domain authority, publishing history, and fact-checking records. By evaluating the reputation and trustworthiness of news sources, these algorithms can identify unreliable sources known for spreading rumors.
In the source-based approach, we first consider the case in which the rumor is generated as a news item.  
In this case, the source may be a news author. 
For example, Sitaula \etal \cite{Sitaula2020} revealed that when disseminating news, disseminating it along with the author's information was more helpful in finding actual fake news. This could be interpreted as a case where the credibility of the author serving as the source was evaluated together. For example, the number of authors or historical news could serve as features. 
Luvembe \etal \cite{Luvembe2023} proposed a deep normalized attention-based mechanism for the enriched extraction of dual emotional features such as 
publisher emotions and social emotions. They used a stacked BiGRU to extract the dual emotional features and suggested using an adaptive genetic weight update random forest (AGWu-RF) to improve the accuracy of fake news detection.
Yuan \etal \cite{Yuan2020} proposed a new structure-aware multihead attention network (SMAN) that combined news content, publishing, and reposting relations of publishers and users for fake news detection. For this, they focused on the credibility of publishers and users to quickly detect fake news.

\begin{algorithm}[t!]
 \caption{Rumor Detection Method}
\label{alg:rumor}
{\small

 \KwIn{Rumor task $T$} 
 \KwOut{Label of $T$}

\smallskip

Set the label of the task $T$, $L(T) =0$;

Perform rumor detection algorithm $f$ for a given task $T$;
\smallskip

 \If{$T$ is regarded as a rumor}{
   $f(T) =0$
   } \ElseIf{ $T$ is not regarded as a rumor}{
    $f(T) =1$}
    
    $L(T)\leftarrow f(T)$;
    
 Return $L(T)$\;
}
\end{algorithm}

User-based rumor detection leverages insights from user-behavior analysis, credibility assessment, influence analysis, community detection, sentiment analysis, and bot detection to identify and mitigate the spread of rumors on social-media platforms. By focusing on individual users and their interactions within the network, these approaches provide targeted interventions to combat misinformation and foster a more trustworthy online environment.
It has been reported that \cite{Ramalingam2018} social robots, sybil accounts, fake profiles, fake accounts, etc., are
generally malicious users of social media and are more likely
to spread fake news. 
Rostami \etal \cite{Rostami2020} demonstrated a feature selection process, which is presented below.
Key processes of machine-based fake-account detection
learning. The way how to select a versatile hybrid feature is 
Used for feature selection
Kudugunta \etal \cite{Kudugunta2018} designed a deep neural network based on contextual long short-term memory (Contextual-LSTM) architecture.
This model used both content and metadata to detect bots at the tweet level. Further, they introduced a synthetic minority oversampling technique that could improve the existing datasets by generating additional labeled examples. 
Bazmi \etal \cite{Bazmi2023} proposed a multiview co-attention network (MVCAN), which jointly considered the latent topic-specific credibility of users and news sources for rumor detection. The core idea was to encode the biased viewpoint of a news article, the user's bias, and the partisan bias of the news source into vectors.
Gao \etal \cite{Gao2020} proposed a novel Sybil detection method using the content, an end-to-end classification model. The designed model included the self-normalizing CNN and bidirectional self-normalizing LSTM network (bi-SN-LSTM) and a dense layer with softmax classifier.
Zhang \etal \cite{Zhang2018} introduced a novel automatic fake-news credibility inference model, namely FAKE-DETECTOR, which
extracted a set of explicit and implicit features
from the text information. 
Qureshi \etal \cite{Qureshi2021} analyzed fake new messages caused by COVID-19 on Twitter using a source-based algorithm and showed high classification performance. 
Khubaib \etal \cite{Khubaib2022} also demonstrated the performance of source-based algorithms through machine learning techniques such as Decision Tree and K-Nearest Neighbor (KNN) in response to misinformation spreading on social media.

\begin{table*}[t!]
\renewcommand{\arraystretch}{1.3}
\caption{Comparative study of selected papers.} \label{tab:rumordp} \centering \begin{tabular}{p{2cm}|p{2cm}|p{2cm}|p{2cm}|p{2cm}|p{2cm}}
\hline      & \textbf{Data Set} &  \textbf{Model} & \textbf{Algorithm}  & \textbf{Metric} & \textbf{Evaluation}  \\
\hline \textbf{DEAP-FAKED \cite{Mayank2022}} & Kaggle Fake News & Knowledge Graph-based & Deep-Faked & F1 score & 0.8926 \\
\hline  \textbf{CompareNet \cite{Hu2021}}  & Satirical and Legitimate News Database& Knowledge Graph-based & CompareNet &F1 score & 0.8912 \\ 
 \hline
 \textbf{Propagation2Vec \cite{Silva2021} }  &PolitiFact and GossipCop. & Cascade-based& Propagation2Vec & F1 score& 0.893\\
  \hline
   \textbf{Contextual-LSTM \cite{Kudugunta2018}}  & Social spambots& User-based & LSTM &Accuracy & 0.9633 \\
  \hline
  \textbf{VRoC\cite{Cheng2020}}  & PHEME & Machine learning & VRoC & F1 score& 0.876 \\
  \hline
  \textbf{Bi-GCN \cite{Bian2020}} & Weibo, Twitter & GCN & Bi-GCN &Accuracy &0.961  \\
  \hline
    \textbf{Graph Neural Network \cite{Liu2024}} & Twitter & GNN& REPORT &F1 score&0.938  \\
  \hline
  \textbf{MultiModal \cite{Peng2023}} & Weibo, Twitter & Deep Learning & MRML &F1 score&0.901  \\
  \hline
  \textbf{$T^3$ RD \cite{Zhang2024}} & Weibo, Twitter-COVID19 & Test Time Training & $T^3$ RD &F1 score&0.917  \\
  \hline
 \end{tabular} \end{table*}

\smallskip
\subsubsection{Hybrid and machine learning-based approaches}

Hybrid rumor-detection algorithms combine content-based, propagation-based, and source-based analysis techniques to leverage the complementary strengths of all approaches. By integrating textual features with the network structure and dynamics, these algorithms can achieve higher accuracy for identifying and classifying rumors.
Baly \etal \cite{Baly2018} conducted a study to find biased news or articles using both source-based and content-based methods. They collected various data from Twitter, Wiki, and the web and implemented it based on fact-checking technology.
Fetterly \etal \cite{Fetterly2005} conducted a study to detect spam floating around a webpage. For this purpose, an algorithm was proposed to classify it based on web sources and content.
Ntoulas \etal \cite{Ntoulas2006} investigated web spam, which is the injection of artificially created pages into the web to influence the results from search engines. They researched to detect such spam using various previous algorithms efficiently.
Benjamin \etal \cite{Benjamin2019} 
proposed various content-sharing techniques based on the source. For this purpose, research was conducted to find fake news by constructing a graph.
Abernethy \etal \cite{Abernethy2010} conducted a study that applied graph-regularization techniques for web-spam detection and analyzed them with respect to source-based methods.

Recently, research has been actively conducted to detect rumors by incorporating machine learning technologies, including NLP.
Mingxi \etal \cite{Cheng2020} designed VRoC, which is a tweet-level variational autoencoder-based state-of-art rumor detection system.  Mingxi \etal \cite{Cheng2021} also proposed a Generative Adversarial Networks (GAN)-based layered model for rumor detection with explanations. Their model provides explanations in rumor detection based on tweet-level texts only without referring to a verified news database.
Lao \etal \cite{Lao2021} considered a novel model named, Rumor Detection with Field of Linear and Non-Linear Propagation (RDLNP) by using of claim content, social context and temporal information.
Mingxi \etal \cite{Mingxi2021} analyze the spread trends of COVID-19 misinformation and found that a log-normal distribution well models the statics of the misinformation. Bian \etal \cite{Bian2020} proposed a novel Bi-Directional Graph Convolutional Networks (Bi-GCN), to explore both characteristics by operating on both top-down and bottom-up propagation of rumors. 
Somya \etal \cite{Somya2020} considered a hybrid approach that leverages the capabilities of various machine learning algorithms called a genetic algorithm to separate spammer and non-spammer contents and account.

 \vspace{-0.3cm}
\subsection{Evaluation Metrics}
As performance metrics, many studies considered (i) recall, (ii) precision, (iii) accuracy, and (iv) F1 score \cite{Samah2018,Li2019}. To formally explain these metrics, true positive (TP) is the number of correctly predicted positive responses, that is, the actual response is yes, and the predicted response is also yes; true negative (TN) indicates the number of correctly predicted negative responses, that is, the actual response is no, and the predicted response is also no; false positive (FP) is when the actual response is no, but the predicted response is yes; false negative (FN) is when the actual response is yes, but the predicted response is no. The three metrics are described as follows:

\smallskip
\begin{enumerate}[(i)]

\item Precision: Precision is the ratio of correctly predicted positive responses to the total predicted positive responses.
\begin{equation}\label{eqn:deg}
\begin{aligned}
Precision = \frac{TP}{TP+FP}
\end{aligned}
\end{equation}

\smallskip
\item Recall: Recall is the ratio of correctly predicted positive responses to all answers in the actual class of responses.
\begin{equation}\label{eqn:clo}
\begin{aligned}
Recall = \frac{TP}{TP+FN}
\end{aligned}
\end{equation}

\smallskip
\item Accuracy: Accuracy is the ratio of correctly predicted responses to the total number of responses.

\begin{equation}\label{eqn:bet}
\begin{aligned}
Accuracy = \frac{TP+TN}{TP+FP+FN+TN}
\end{aligned}
\end{equation}
Accuracy is a good measure when the values of false positives and false negatives of the data sets are almost the same

\item F1 Score: This metric is a weighted average of precision and recall. Therefore, this score considers both false positives and false negatives as follows:
\begin{equation}\label{eqn:bet}
\begin{aligned}
F1~score = \frac{2 \times precision \times recall}{precision + recall}
\end{aligned}
\end{equation}

The F1 score is usually more useful than accuracy when the values of false positives and false negatives of the data sets are quite different.

\end{enumerate}
Based on this, we summarized some empirical results in Table~\ref{tab:rumordp}.

\begin{table*}[t!]
\renewcommand{\arraystretch}{1.3}
\caption{Taxonomy of rumor source detection algorithms.} \label{tab:rumorsd} \centering \begin{tabular}{p{2cm}|p{6cm}|p{6cm}}
\hline      & ~~~~~~~~~~~~~~~~~~~~~~~~\textbf{Single source} & ~~~~~~~~~~~~~~~~~~~~ \textbf{Multiple sources} \\
\hline \textbf{Complete Snapshot} & Rumor center \cite{shah2010,shah2012,shah2010tit}, Jordan center \cite{zhu2013,zhu2014}, Multi-snapshots \cite{Zhang2014}, Prior set \cite{dong2013}, Decaying diffusion rate \cite{Choi2019}, Querying \cite{Choi2017,Choi2018}, Set estimator \cite{Bubeck14,Khim14}, Random growing graph \cite{Fuchs2015},   Short-Fat Tree \cite{zhu2016}, Federated learning \cite{Wang2024} & Minimum description length \cite{ICDM17}, Different diffusion time \cite{JI17}, Neighborhood entropy \cite{Liu2022}, Graph convolutional networks \cite{Dong2019}, Effective distance \cite{Jiang2019}, Joint rumor center \cite{Wang2015}\\
\hline  \textbf{Partial Snapshot}  & Infection path \cite{Leng2014}, Reverse dissemination \cite{Jiang19}, NETFILL \cite{Shashidhar2015}, Probabilistic sampling \cite{Nikhil2013}, Noisy observation \cite{Fabrizio2014}, Infection likelihood \cite{Alexandru2019}& Messenger selection \cite{Hu18}, Jordan Cover \cite{Kai2017} Pearson based \cite{Wang2023}, Anomalous sources detection \cite{Zhang2016}, Minimum SRS \cite{Zhang2017}\\
 \hline
 \textbf{Sensor Monitoring Snapshot}  &  General framework\cite{Brunella2019} Gaussian Method\cite{Pinto2012} Monte Carlo \cite{Agaskar2013} Belief propagation \cite{Altarelli2014} Four metric estimator \cite{Seo2012}, Reverse dissemination \cite{Jiang19}, ROSE \cite{Ravi2024}& Fact checkers \cite{Chen2021}, K-MLE \cite{Huanga2024}\\
  \hline
 \end{tabular} \end{table*}

  \begin{table*}[t!]
\renewcommand{\arraystretch}{1.3}
\caption{Summary of rumor source detection algorithms.} \label{tab:rumorsds} \centering \begin{tabular}{p{2cm}|p{2cm}|p{1cm}|p{5cm}|p{5cm}}
\hline    & ~~\textbf{Ref}  & ~~\textbf{$\#$ of sources} & ~~\textbf{Pros} & ~~\textbf{Cons}\\
\hline \textbf{Complete Snapshot} & Rumor center \cite{shah2010,shah2012,shah2010tit} & Single & Theoretical analysis of the actual detection probability in regular tree and derivation of a new graphic centrality concept & Simple setting such as homogeneous diffusion rate, regular tree, etc. \\  
\cline{2-2} \cline{3-2} \cline{4-2} \cline{5-2} & Jordan center \cite{zhu2013,zhu2014}& Single & Theoretical analysis and MLE derivation considering infection time & 
Complex formulas and theoretical limitations \\
\cline{2-2} \cline{3-2} \cline{4-2} \cline{5-2} & Multi-snapshots \cite{Zhang2014}& Single & Theoretical proof that detection probability increases when multiple snapshots exist & 
Simple extension of Rumor center \\
\cline{2-2} \cline{3-2} \cline{4-2} \cline{5-2} & Prior set \cite{dong2013}& Single & Theoretical proof that detection probability increases when prior culprit set is given & 
Simple setting and naive extension of Rumor center \\
\cline{2-2} \cline{3-2} \cline{4-2} \cline{5-2} & Querying \cite{Choi2017,Choi2018} & Single & Increase detection probability by appropriately using additional querying information & 
Assume that all nodes have the same probability of telling the truth \\
\cline{2-2} \cline{3-2} \cline{4-2} \cline{5-2} & Minimum description length \cite{ICDM17} & Multiple & A new approach of eigenvector estimator to infection graphs & 
Lack of accurate detection probability performance metrics \\
\cline{2-2} \cline{3-2} \cline{4-2} \cline{5-2} & Different diffusion time \cite{JI17} & Multiple & Consider practical situations where multiple rumor sources spread information at different times & 
Assumption that the number of source nodes is already known \\
\cline{2-2} \cline{3-2} \cline{4-2} \cline{5-2} & Graph convolutional networks \cite{Dong2019} &  Multiple & Efficient algorithm development without assumptions about the underlying propagation model & 
Lack of explanation as to why the model performs well \\
\hline  \textbf{Partial Snapshot}  & Infection path \cite{Leng2014} & Single & Consider practical situations where the state of an arbitrary node may be hidden & Homogeneous propagation rate\\
\cline{2-2} \cline{3-2} \cline{4-2} \cline{5-2} & Reverse dissemination \cite{Jiang19} & Single & Consider graph topology that may change over time & 
Relatively short rumor spreading time\\
\cline{2-2} \cline{3-2} \cline{4-2} \cline{5-2} & NETFILL \cite{Shashidhar2015} & Single & Perform recovery operation on missing (unobserved) node & 
Simple SI propagation with homogeneous rate\\
\cline{2-2} \cline{3-2} \cline{4-2} \cline{5-2} & Probabilistic sampling \cite{Nikhil2013} & Single  & Theoretical analysis of detection performance using sampling probability of whether a node heard a rumor & 
Analyzing only regular tree structure\\
\cline{2-2} \cline{3-2} \cline{4-2} \cline{5-2} & Jordan Cover \cite{Kai2017} & Multiple  & Theoretical analysis of heterogeneous diffusion rate and detection probability in ER random graph & 
Assumption that the number of source nodes is already known\\
\cline{2-2} \cline{3-2} \cline{4-2} \cline{5-2} & Anomalous sources detection \cite{Zhang2016} & Multiple  & Analysis performed without assumptions about the number of rumor sources and settings for initial propagation time & 
No direct performance comparison with other algorithms\\
\cline{2-2} \cline{3-2} \cline{4-2} \cline{5-2} & Minimum SRS \cite{Zhang2017} & Multiple  & Efficient graph theoretical analysis of infection graphs & 
No numerical or simulation experiments on actual data\\
 \hline
 \textbf{Sensor Monitoring Snapshot}  & General framework\cite{Brunella2019} & Single & Online and offline source detection through static and dynamic sensors & There are not many existing methods used for performance comparison\\
 \cline{2-2} \cline{3-2} \cline{4-2} \cline{5-2} & Gaussian Method\cite{Pinto2012} & Single  & Consider a more general situation where transmission delay follows Gaussian distribution and derive an efficient algorithm & 
There is only analysis of tree-like structures\\
 \cline{2-2} \cline{3-2} \cline{4-2} \cline{5-2} & Belief propagation \cite{Altarelli2014} & Single  & High performance due to BP algorithm that works well in tree structure & 
Lack of experimentation and analysis in general graphs\\
 \cline{2-2} \cline{3-2} \cline{4-2} \cline{5-2} & Four metric estimator \cite{Seo2012} & Single  & Efficient algorithm design based on the simple assumption that the source will be located close to the sensor node & 
Performance comparison with relatively simple graph centrality\\
 \cline{2-2} \cline{3-2} \cline{4-2} \cline{5-2} & Fact checkers \cite{Chen2021} & Multiple  & Consider the more complex SEIR model rather than the SI or IC model & 
There is only a performance comparison with rumor center\\
 \cline{2-2} \cline{3-2} \cline{4-2} \cline{5-2} & K-MLE \cite{Huanga2024} & Multiple  & Proposal of source detection problem considering temporal dynamics of community structure in network & 
Lack of algorithms to compare the performance of the proposed algorithm\\
  \hline
 \end{tabular} \end{table*}

\subsection{Rumor Source Detection Methods and Algorithms}
A rumor-source--detection algorithm uses snapshot information or information obtained from sensors when a rumor spreads in the network to find the node that first spread the rumor. We also describe a simple pseudocode in Algorithm~\ref{alg:rumors}. This is an inference algorithm that estimates the source based on snapshot data and graph information, and when it spreads stochastically, it becomes a combination problem that requires the calculation of a large number of cases. 
To summarize the approaches to this problem, we first have divided and organized it as in Table \ref{tab:rumorsd}, according to the number of rumor sources and type of snapshot data and give some detailed description in Table \ref{tab:rumorsds}, respectively.

\smallskip
\subsubsection{Single source detection algorithms $(|S^*|=1)$}
When a rumor spreads in a network, if all information about which node was infected is given, the scenario is called a complete snapshot. For a given complete snapshot information, 
Shah \etal \cite{shah2010,shah2012,shah2010tit} considered the source-detection problem of the rumor spreading over the network as a primary study. They introduced a metric
called rumor centrality, a simple graphical
metric for a given propagation snapshot. 
A rumor center, which had the maximum rumor centrality, was introduced as the MLE.
They proved that rumor centrality implied the likelihood
function when the underlying network was a regular tree and that the diffusion followed the SI model, which was
extended to a random graph network in \cite{shah2012}. Zhu \etal
\cite{zhu2013} addressed the source-detection problem under the SIR diffusion model and considered a sample path approach to
solve the problem and prove that the likelihood estimator based on the sample path was exactly the Jordan center over the infection graph. Further, they extended it to the case of limited observations \cite{zhu2014}. In addition to this, several works have been introduced to boost the detection probability using some side information.
Wang \etal \cite{Zhang2014} showed that multiple different
rumor snapshots from a single source could significantly increase the detection
probability. Dong \etal \cite{dong2013} showed that if prior information about the source was available, then the detection probability would be increased dramatically if the maximum posterior estimator (MAPE) was used.

Choi \etal \cite{Choi2019} showed that when an antirumor initiator existed in the network, and the distance distribution between the anti-rumor source and rumor source was given, 
the rumor source could be found using the MAPE. Choi \etal \cite{Choi2017,Choi2018} studied the effects of
querying to find the source and demonstrated the necessary and sufficient number of queries 
to achieve a target detection probability.
The authors in \cite{Bubeck14,Khim14} introduced the notion of
  set estimation and provided analytical results of the detection
performance.
For a dynamic and randomly increasing network, Fuchs \etal \cite{Fuchs2015} considered rumor propagation in several randomly grown graphs such as binary search trees, recursive trees, and plane-oriented recursive trees. They used rumor centrality to derive an analytical formula for the detection probability of these simple, randomly increasing trees. Furthermore, they briefly discussed the central rumor about unordered trees.
Zhu \etal \cite{zhu2016} considered the source-detection problem in ER random graphs. They proposed the short-fat tree (SFT) algorithm, a new source-localization algorithm for the IC diffusion model. This algorithm selected a node such that the depth of the BFS tree from that node was the minimum while the number of leaf nodes was the maximum. They obtained an analysis of the basic limitations by considering the duration of infection.

When a rumor spreads in a network, the situation in which information about all infected nodes is not given, and some cannot be confirmed is called a partial snapshot observation. Many studies have been conducted to find sources of rumor in such practical environments.
Luo \etal \cite{Leng2014} considered the problem of estimating an infection
source for the SI model, in which not all infected
nodes could be observed. Jiang \etal
\cite{Jiang19} considered a time-varying topology to infer the rumor source. To do this, they reduced the time-varying networks to a series of static networks by introducing a time-integrating window. Then, instead of inspecting every individual using traditional techniques, they adopted a reverse dissemination strategy to specify a set of suspects of the real rumor source. 
Shashidhar \etal \cite{Shashidhar2015} considered noisy infections that could occur when rumors spread in the network. Their study proposed a technology that accurately inferred the source of the rumor by recovering missing infections. To this end, they developed an algorithm called NETFILL, a new technique for solving the problem of the Minimum Description Length (MDL) principle.
Nikhil \etal \cite{Nikhil2013} modeled a situation in which each node reported its infection status with probability $p\in (0,1)$ when a rumor spread in the network. They used an estimator of rumor centrality in the SI model to study how many source detections occurred when $p$ satisfied certain conditions.
Fabrizio \etal \cite{Fabrizio2014} proposed an approach for cases where all information was not obtained, owing to noise, while observing snapshots in the rumor-propagation model of the SIR model. 
To this end, they constructed a factor graph and solved the source-detection problem with the BP algorithm, which worked well in a tree-like structure.
Alexandru \etal \cite{Alexandru2019} introduced a new method to estimate the source of multiple rumors in an arbitrary network topology with partial observations of the network nodes. Under the SI model, they considered two mathematical discrete-time models that could represent the propagation process well with low complexity.

In some cases, sensors are mounted on specific nodes and observed to detect rumors spreading in the network. In this case, a method of estimating the source of the rumor using the data observed by the sensors was proposed.
Spinelli \etal \cite{Brunella2019} broadly divided the sensors in the network into static sensors and dynamic sensors and conducted source estimation by considering a scenario in which specific sensors could be added arbitrarily. In addition, they conducted research by creating a general framework that considered both online detection that occurred during propagation and offline source detection that occurred after propagation. They discovered that even a small amount of dynamic sensor use could be very helpful in source estimation.
Pinto \etal \cite{Pinto2012} proposed
a Gaussian method for rumor-source estimation that 
assumed a deterministic propagation time for each
edge, which was independent and identically distributed with
Gaussian distribution. They considered tree-like network structures and focused on the BFS tree structure as the phenomenon of rumor spreading in this tree structure is very similar to the BFS tree form. Research was conducted taking this into consideration.
Agaskar \etal \cite{Agaskar2013} proposed a fast Monte Carlo method for source identification in generic networks based on observations from a small set of observers. They considered a heterogeneous SI model in which the probability of infection propagating through the edges connecting all nodes was not the same. In addition, the sensors were observed in a fixed time window. However, as they had only focused on the assumption that information always spread to other nodes along the shortest paths, applying this in real scenarios appeared to be difficult.
Altarelli \etal \cite{Altarelli2014} considered the Bayesian belief-propagation model and obtained a good detection performance for a tree-like network structure. SI and SIR models were considered the rumor-propagation model, and they formed a factor graph to infer this source by considering the belief-propagation algorithm.
Seo \etal \cite{Seo2012} considered a directed network for rumor propagation and suggested a four-metric source estimator to find the source node. Under the SI model, they assumed the sensor node to be positive when it transited from susceptible states
to infected states; otherwise, it was regarded as a negative sensor. They compared various other methods for choosing the sensor locations, such as 
random choosing
(Random), choosing the nodes with high betweenness centrality
values (BC), choosing the nodes with a large number
of incoming edges (NI) and choosing the nodes that were
at least d hops away from each other (Dist)

\smallskip
\subsubsection{Multiple source detection algorithms $(|S^*|>1)$}

Several studies tried to solve the problem of finding multiple sources by appropriate set estimation methods. First, we consider the multiple-source detection problem under the complete-snapshot observation scenario. 
Prakash \etal \cite{ICDM17} proposed to employ the minimum description length (MDL) principle to identify the best set of sources and virus propagation ripple, which describes the infected graph most succinctly.
They proposed an efficient algorithm to identify likely sets of source nodes
given a snapshot and showed that it could optimize the
virus propagation ripple in a principled way by maximizing the likelihood. 
Ji \etal \cite{JI17} developed a theoretical framework to estimate rumor sources, given an observation of
the infection graph and the number of rumor sources.
Liu \etal\cite{Liu2022} conducted a study to find the sources of the entire infection graph by appropriately calculating the entropy of each node's neighboring nodes from among the information spread by the network's rumor. 
To this end, they defined the concepts of infection intensity and infection intensity entropy and developed an algorithm to estimate the locations of multiple sources by calculating them for each node.
Dong \etal \cite{Dong2019} developed a deep-learning--based source-estimation algorithm called graph-convolutional-network--based source identification (GSNSI). This algorithm calculated the spread of labels in the graph using a graph convolutional network and compared it with the values of the surrounding nodes to find the source node. 
Jiang \etal \cite{Jiang2019} proposed an algorithm that found multiple sources based on a metric called the effective distance based on the infected nodes in the network. They obtained results that worked relatively well in various propagation models.
Wang \etal \cite{Wang2015} researched how to find the source of a rumor when it spread multiple times from multiple sources. 
They developed an expanded rumor centrality that could be applied to multiple snapshots and used it as a new estimator. They theoretically revealed in a tree structure that source estimation could be done very well, even with a small amount of snapshot information.

Next, we will see how the multiple-source detection problem can be solved when the complete snapshot information is not given, and only a part of it is given.
Hu \etal \cite{Hu18} considered the problem of locating multiple diffusion sources in time-varying networks. They developed a general framework to locate diffusion sources in time-varying networks based solely on sparse data from a small set of messenger nodes. They found that large-degree nodes produced more valuable information than small-degree nodes, contrasting static networks. By choosing large-degree nodes as the messengers, they also found that sparse observations from a few such nodes were often sufficient for any number of diffusion sources to be located for a variety of models and empirical networks.
Zhu \etal \cite{Kai2017} proposed a new source-localization algorithm named optimal Jordan cover (OJC). Using a candidate-selection algorithm, the algorithm first extracted a subgraph that selected source candidates based on the number of observed infected nodes in their neighborhoods. Then, in the extracted subgraph, OJC found a set of nodes that covered all the observed infected nodes within a minimum radius. Considering the heterogeneous SIR diffusion in the ER random graph, they proved that OJC could locate all sources with unit probability asymptotically with partial observations. 
Wang \etal \cite{Wang2023} proposed a multisource detection
algorithm for different propagation dynamics for given partial observations. For this, they designed an algorithm that calculated a personal correlation between the information times of the nodes and the geodesic distance between the nodes and sources. They extracted a new centrality of infection graph and obtained the detection performance. 
Zhang \etal \cite{Zhang2016} proposed a multiple-source detection problem given only a small subset of network nodes. For this, they used a novel regression learning model by jointly solving five challenges: an unknown number of source nodes, a few activated detectors, an unknown initial propagation time, an uncertain propagation path, and an uncertain propagation delay. They observed the performances bound by theoretical analysis. 
Zhang \etal \cite{Zhang2017} considered the multiple-source--detection problem using a set resolution set (SRS) problem from a deterministic point of view. For this, they proposed a polynomial-time greedy algorithm for finding
a minimum SRS in a general network with $O(\ln n)$ performance ratio, where $n$ is the number of nodes in the network. They showed that their proposed method was robust and efficient for any propagation model.

\begin{algorithm}[t!]
 \caption{Rumor Source Detection Method}
\label{alg:rumors}
{\small

 \KwIn{Graph $G$, Infection snapshot $G_I$, Infection probability $P$, Propagation model $M$} 
 \KwOut{Set of estimated rumor sources $\widehat{S}$}

\smallskip

 $\widehat{S} = \emptyset$;

\For{$v \in V$}{
Perform a rumor source detection algorithm using $(G,G_I, P, M)$;

\smallskip

 \If{$v$ is an estimated rumor source}{
 $\widehat{S}\leftarrow \widehat{S} \cup \{v\}$;
   } \ElseIf{$v$ is not an estimated rumor source}{ $\widehat{S}\leftarrow \widehat{S} $;
    }

}
}
 Return $\widehat{S}$;
\end{algorithm}

Finding the source of rumors spread in the network, based on sensor measurements, is somewhat useful when there is only one rumor source; however, it becomes a difficult problem when there are multiple rumor sources.
Chen \etal \cite{Chen2021} added some ``fact-checkers" over the infectious propagation model for multiple rumor-source detection. For this, they constructed a new propagation model named the SIDR model in social networks and improved the performance by adopting the beam search algorithm. They showed that their proposed algorithm was up to 83\% at the early stage.
Huanga \etal \cite{Huanga2024} proposed a multiple rumor-source detection algorithm based on the placement of sensors for community-based OSNs. They simultaneously conducted research on strategies for efficiently deploying sensors to find the source.
The main idea of the multisource estimation method is as follows:
After an initial assessment of the community's potential to spread the word
within the communities, single-source detection is performed. The algorithm integrates the observations from multiple sensors to identify the
potential communities that harbor rumor sources. Subsequently, it employs the maximum likelihood estimation method within these prospective communities to estimate the sources.

\subsection{Joint Detection Algorithms}
Joint inference algorithms for rumors and their sources aim to simultaneously identify a rumor's existence and its origin within a network or community. A sample pseudocode is described in Algorithm ~\ref{alg:joint}. This is a very efficient way to infer two things at the same time; however, the approach is complex and has not been considered much to date. The research conducted so far is summarized below.

\smallskip
\subsubsection{Source reliability-based approach \cite{Qu2020}}
This study proposed a joint inference algorithm that estimated a source that had been disseminating information in social networks, measured the credibility and reliability of the source, and simultaneously discerned whether the disseminated information was a rumor or true. For this, they proposed a framework named
SourceCR, which consisted of two processes: $(1)$ credibility-reliability
training for truth/rumor inference and $(2)$ division-querying for
source inference. In the first process, they used the expectation-maximization (EM) algorithm to maximize the likelihood function of the user's opinion for each piece of information. 
The first process used the reliability of the source obtained from the second process as input.
In the second process, they split the network into two subnetworks for each claim, which they labeled as correct or incorrect based on the claim confidence evaluation returned from the previous module. Furthermore, in each subnetwork, they choose to query some users according to their reliabilities estimated by the previous process, and based on their answers, they infer the source of a theoretically guaranteed lower bound within a given budget.
Accurate measurements on the claim credibility and user reliability were guaranteed because of the use of an optimal solution derived from the algorithm.
These two processes operated iteratively, with one process using the parameters output by the other process in the previous iteration as inputs for the next iteration.

\smallskip
\subsubsection{Number-of-sources--based approach \cite{Seo2012}}
In this approach, the authors focused on two issues related to mitigating false claims in social networks. First, they studied the problem of identifying the source of a rumor in the absence of complete source information about rumor propagation. 
Second, they studied how to distinguish rumors (false claims) from fact (true information). This method assumed that rumors originated from a small number of sources, whereas truthful information was simultaneously observed and originated by many unrelated individuals. Their approach relied on utilizing network monitors, which were individuals (on their social networks) who had agreed to inform when they had heard certain information. However, they would not disclose who had provided them with the information or when they learned of it.

Although the joint problem may be more helpful in solving each problem, much research has been conducted separately on the above two problems to date. The authors of \cite{Seo2012} conducted a study that considered the above two studies simultaneously for the first time. 
In addition, a study was recently conducted by the authors of \cite{Qu2020} to infer both the rumor and its source simultaneously by combining the two.

\subsection{Evaluation Metrics}
Many evaluation metrics are used to evaluate the performance of algorithms in rumor source detection. To summarize, it is as follows.

\smallskip
\begin{enumerate}[(i)]

\item Detection probability (detection rate): This is the most widely used performance measure used in \cite{shah2010,shah2012,shah2010tit}. It is the ratio between the number of detections and the number of simulations
\begin{equation}\label{eqn:deg}
\begin{aligned}
P(detection) = \frac{\# ~\text{of detections}}{\# ~\text{of diffusions}},
\end{aligned}
\end{equation}
Here, $\#$ of diffusions indicates the number of simulations of rumor propagation for estimating the sources and $\#$ of detections means that the number of events $\{\widehat{S}=S^*\}$, \ie the estimator detects the sources. 

\smallskip
\item F1-score: This measured by precision and recall which is described as 
\begin{equation}\label{eqn:clo}
\begin{aligned}
F1 score = \frac{2 \times precision \times recall}{precision + recall},
\end{aligned}
\end{equation}
where 
\begin{equation}
\begin{aligned}
precision = \frac{|\text{Detected sources}|}{|\text{Estimated sources}|},
\end{aligned}
\end{equation}
and 
\begin{equation}
\begin{aligned}
recall = \frac{|\text{Detected sources}|}{|\text{True sources}|},
\end{aligned}
\end{equation}

\smallskip
\item Distance error: The distance error is referred as the shortest distance of hops between the true source and estimated source found by an algorithm, \ie

\begin{equation}
\begin{aligned}
Err_{dist} = d(\widehat{S},S^*)
\end{aligned}
\end{equation}
If the source is single, then the distance is the shortest path between the estimator and the true source. However, for the multiple sources, additional calculations are required to measure the distance between the two sets.

\smallskip
\item Rank: It means the location of the actual source among the nodes in decreasing order by calculating the score among the nodes where the rumor spread in the network based on a specific standard. The ranking measure is well suited for identifying a small group of nodes among which the source node is present \cite{Sushila2018}.

\end{enumerate}
Based on this, we summarized some empirical results in Table~\ref{tab:rumorsdp}.

\begin{algorithm}[t!]
 \caption{Joint Rumor and Rumor Source Detection Method}
\label{alg:joint}
{\small

 \KwIn{Rumor task $T$, Graph $G$, Infection snapshot $G_I$, Infection probability $P$, Propagation model $M$, Node reliability $R$} 
 \KwOut{Task label and set of estimated rumor sources ($L(T), \widehat{S}$)}

\smallskip

 $\widehat{S} = \emptyset$ and $L(T)=0$;

\For{$v \in V$}{
Perform a rumor source detection algorithm using $(G,G_I, P, M)$;

\smallskip

 \If{$v$ is an estimated rumor source}{
 $\widehat{S}\leftarrow \widehat{S} \cup \{v\}$;
   } \ElseIf{$v$ is not an estimated rumor source}{ $\widehat{S}\leftarrow \widehat{S} $;
    }
    }
    Perform rumor detection algorithm $f$ with the node reliability of $\widehat{S}$ or $|\widehat{S}|$ for a given task $T$;
\smallskip

 \If{$T$ is regarded as a rumor}{
   $f(T) =0$
   } \ElseIf{ $T$ is regarded as a rumor}{
    $f(T) =1$}
    
    $L(T)\leftarrow f(T)$;
    
 Return $(L(T), \widehat{S})$;
 }
\end{algorithm}

\smallskip

\vspace{-0.5cm}
\section{Hiding the Rumors and their Sources}
\label{sec:hiding}

\begin{figure}[t!]
\begin{center} \centering
\includegraphics[width=1\linewidth]{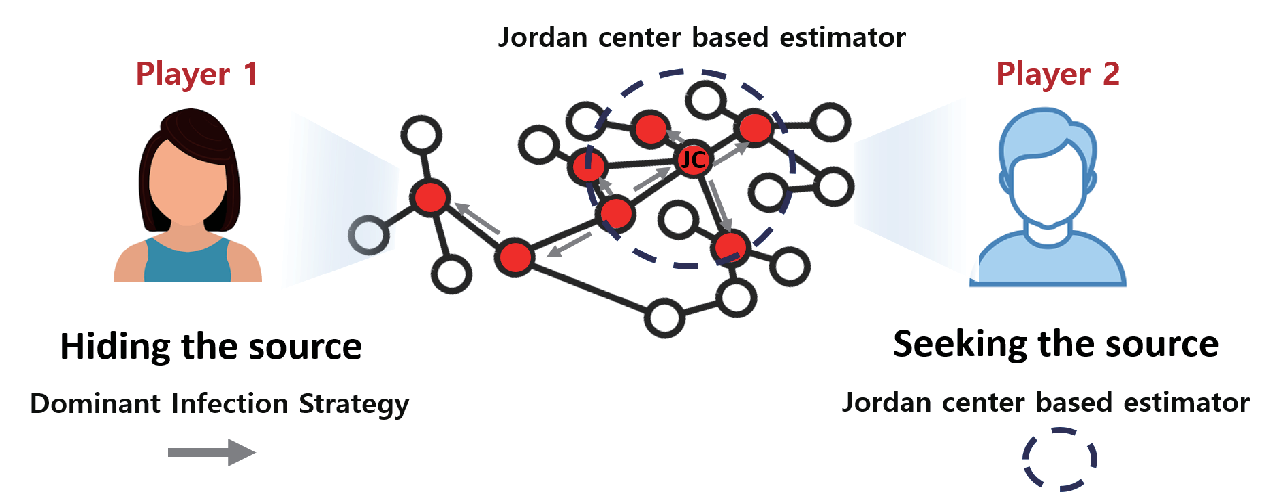}
\caption{A hide-and-seek game of rumor sources \cite{Luo16}. In the game, one player uses a strategy to control the rumor propagation rate at each edge to hide the rumor source. To find the source of the rumor in such a situation, another player establishes an estimation set centered on the well-known Jordan Center (JC) and allows the source to enter it. }
\label{fig:game}
\end{center}
 \vspace{-0.6cm}
\end{figure}

Hiding rumors and their sources can have various implications and motivations, both positive and negative. We summarize this problem as follows.

\begin{table*}[t!]
\renewcommand{\arraystretch}{1.3}
\caption{Comparative study of selected papers. ($N$ = the number of infected nodes in the network, $k$ = number of diffusion snapshots  $m$ = number of sources, $p$ = parameter of model}, respectively.) \label{tab:rumorsdp} \centering \begin{tabular}{p{2cm}|p{2cm}|p{2cm}|p{2cm}|p{2cm}|p{2cm}|p{2cm}}
\hline      & \textbf{Snapshot} &  \textbf{Diffusion Model} & \textbf{Estimator}  & \textbf{Real(Synthetic) World Data} & \textbf{Evaluation}  & \textbf{Complexity}\\
\hline \textbf{Rumor center \cite{shah2010,shah2012,shah2010tit}} & Complete & SI & Ruomor center & Power Grid Network & Detection Probability: 4-5 \% & $O(N)$\\
\hline  \textbf{Jordan center \cite{zhu2013,zhu2014}}  & Complete& SIR & Jordan center &Wikipedia & Distance Error: 1-2 hops & -\\ 
 \hline
 \textbf{NETFILL \cite{Shashidhar2015}}  & Partial & SI& Minimum Description Length & Crid-Con& Precision: 0.7 &$O(|E|+|V|)$\\
  \hline
   \textbf{Joint rumor center \cite{Wang2015}}  & Complete& SI & MLE &Newmans scientific
collaboration network & Detection probability: 20\% ($k=5$)&$O(kN)$\\
  \hline
  \textbf{Jordan Cover \cite{Kai2017}}  & Partial & SIR & Optimal Jordan Cover & Power Grid Network& Detection Probability: 5-8\% &$O((m+|V|)N)$\\
  \hline
  \textbf{Reverse dissemination \cite{Jiang19}} & Sensor monitor &SIR & Reverse Dissemination &Facebook network &Distance error: 1-3 hops  & - \\
  \hline
    \textbf{Federated learning \cite{Wang2024}} & Complete &SI, SIR, IC & Deep and Federated learning &Karate&F1 score: 0.1231 & $O(|V|p)$\\
  \hline
   \textbf{Incomplete observer \cite{Ravi2024}} & Sensor monitor &SI, SIR, SIS, IC & Path-based estimator &Facebook network  &Distance error: 1.39 hops & $O((\log N) * N^2)$ \\
  \hline
 \end{tabular} \end{table*}

\vspace{-0.3cm}
\subsection{Hiding the Rumor}

In some cases, individuals may want to conceal their involvement in spreading rumors to protect their privacy and avoid potential repercussions or harassment.
Concealing the identity of the rumor can prevent retaliation or backlash from individuals or groups affected by the dissemination of false information.
Protecting anonymity may encourage individuals to share sensitive information or express dissenting opinions without any fear of judgment or reprisal.
There have been several works on hiding some information on social networks. 
Szczypiorski \etal \cite{Szczypiorski2016} proposed an approach of stegHash to hide some information in message transmission.
It was based on using hashtags in various social networks to create an invisible chain of various multimedia objects, such as images, videos, or songs, with a hidden message embedded within. The initial set of hashtags became the basis for building an index as a unique pointer to these files. For a set of hashtags containing $n$ elements, there were $n$ factorial permutations, with each instance producing a separate and unique index for a piece of data. The system consisted of a hidden initial set of hashtags (passwords) and a secret transition generator. It started with an initial set of hashtags, and using a transitive generator, connections between indices were created, which could then sequentially explore the hidden data sent from one to the other. One of the existing ideas for applying StegHash technology was to establish an index system similar to existing classic file systems such as the File Allocation Table (FAT) or New Technology File System (NTFS).
This created a new type of steganographic file system in an area that previously refused to operate on a file system or denied the existence of stored data. Bieniasz \etal \cite{Bieniasz2017} proposed a new method named SocialStegDisc, which was a proof of
concept for the application of the StegHash \cite{Szczypiorski2016} method for new
steganographic file systems.  
Some changes were made to the original environment of StegHash to derive basic concepts from classic file systems, such as the create--read--update--delete operations or the defragmentation process.
The design also led to a time-memory trade-off. This concept has been considered in practice to obtain operational results and proof of accuracy.

\vspace{-0.3cm}
\subsection{Hiding the Rumor Sources}

Hiding the sources of rumors can make it difficult to hold individuals or organizations accountable for spreading false or harmful information, undermining the efforts to combat misinformation and promote transparency.
Moreover, anonymity may embolden individuals to engage in malicious activities, such as spreading rumors or disinformation campaigns, without the fear of consequences.
 As opposed to finding the information source from a given
snapshot of the epidemic, hiding the corresponding source approach has also
been studied. Fanti \etal \cite{Oh2015} first considered this problem and proposed
an {\em adaptive diffusion} (AD) method for the information spreading protocol. 
AD is a propagation method that maintains the source that initially spreads the information as a leaf node in the infection graph of information spread in the network. Information about how far each node is from the leaf is shared among the nodes through message passing, and information is spread to other places to prevent the source node from being located in the center of the infection graph, breaking the balance and Hide in the leaf node.
They showed
that AD was near-optimal for hiding the source, as well as
maximizing the information spreading on the regular tree structures, i.e., the probability of detection
for the MLE when $n$ nodes are infected was close
to $1/n$:
\begin{equation}\label{eqn:ad}
\begin{aligned}
P(\hat{v}_{\tt ml} = s^*|N_T =n)=\frac{1}{n}+o\left(\frac{1}{n}\right),\end{aligned}
\end{equation}
where $N_T$ was the number of infected nodes at time $T$ and $s^*$ was the true source. This was the best source obfuscation that could be achieved by
any protocol, as there are only n candidates for the source
and they are all equally likely.
Fanti \etal \cite{Oh2016} also considered a case for an irregular
tree as the underlying topology and showed that the proposed
diffusion protocol was also near-optimal under some conditions. 
In fact, unlike a regular tree, in the case of an irregular tree, the propagation path is complex, which makes theoretical analysis extremely difficult. To overcome this, the authors characterized the most frequently occurring typical topology in the random tree model and analyzed the concentration of probability of detection based on this. As a result, it was theoretically revealed that even in irregular trees, the detection probability of MLE or MAP was maintained very low in the case of AD.
Fanti \etal \cite{Oh2017} extended the proposed AD model from a network to a general graph with a loop. As a general graph, we specifically considered a two-dimensional grid network and proved the theoretical hiding performance guarantee for AD on this. 
They also provided
simulation results from running AD over an
underlying connected network of 10,000 Facebook users,
described by the Facebook WOSN dataset as a social network. They eliminated
all nodes with fewer than three friends, which left us with a network of 9,502 users.
Simulations confirmed that the proposed AD guaranteed the anonymity of the source location even in a somewhat complex social graph.

\vspace{-0.4cm}
\subsection{Game Theoretic Approach}

Luo \etal \cite{Luo16} considered
a problem wherein an information source tries to hide while
maximizing the spread of its information, whereas the network
adversary seeks the source simultaneously, as depicted in Fig~\ref{fig:game}. They formulated a strategic game in which the network
administrator and infection source were the players. In the game, one player, as the network manager, used a source estimator that could examine all nodes within a given radius from a randomly selected Jordan center of the observed infection graph.
Another player was the infection source, and the strategy was to secure the minimum number of safe hops from the Jordan center by controlling the infection rate for each edge. From the network administrator's perspective, the source could be found through a larger investigation, but in this case, more costs could be incurred. From the rumor source's perspective, the reward of being punished the moment it was detected by the network administrator was considered. Given the margin of safety for the infection source, they believed that the network administrator's best response strategy would be to use the Jordan center as the source estimator.
It was shown that either using or adopting an estimated radius equal to the safety margin could be done.
They derived the conditions for the optimal best response theoretically. They showed that, given an estimated radius to the network manager, the optimal safety margin of the infection source would be one greater than the estimated radius or zero if the underlying network was a tree.
They derived a “dominant infection strategy” (DIS), which was an infection strategy that maximized the number of infected nodes given a safety margin.
The authors then derived the conditions under which the Nash equilibrium of the strategic game existed and theoretically proved that when a Nash equilibrium existed, the network manager's best response would be to use the Jordan central estimator.
It provided a game-theoretic interpretation of the Jordan-centered estimator and also serves as a universally robust estimator.

Overall, the decision to hide rumors and their sources should be carefully considered, taking into account factors such as privacy rights, accountability, and the potential impact on trust and information integrity within online communities. While anonymity can protect individuals from harm in some cases, it also poses risks to the reliability and credibility of information shared online. Finding a balance between privacy and transparency is essential to promote a healthy and trustworthy online information ecosystem.

\section{Discussion and Challenges}
\label{sec:discussion}
In this section, we will describe some fundamental challenges for the three problems and some possible future direction as in Table ~\ref{tab:direction}.

\subsection{Challenges Associated with Rumor Detection}
Rumor detection is associated with several challenges because of the complex nature of misinformation and dynamic environment of social-media platforms. Here are some of the key challenges:

\smallskip
\subsubsection{Data volume and velocity}
Social media generates vast amounts of data in real time, making it challenging to process and analyze information quickly enough to detect rumors as they emerge \cite{Bazzaz2020}. The rapid velocity of information propagation further complicates the task of distinguishing rumors from legitimate content.
Social media platforms generate massive volumes of data in the form of posts, comments, likes, shares, and interactions among users. This sheer volume of data makes it challenging for rumor-detection systems to process and analyze all the information effectively.
Information spreads rapidly on social media, with rumors often circulating within minutes or even seconds of their emergence. This rapid velocity of information propagation means that rumor-detection systems must be able to analyze and respond to new information in real time to prevent the spread of false or misleading information \cite{Liu2023}.


\smallskip
\subsubsection{Variability in rumor characteristics}
Rumors can take various forms, ranging from false claims and conspiracy theories to satire and parody. The variability in rumor characteristics refers to the diverse range of attributes and features exhibited by rumors as they propagate through social networks \cite{Imad2020}. Understanding and identifying these characteristics is essential for effective rumor detection. 
Rumors can cover a wide range of topics, including politics, health, celebrity gossip, natural disasters, and conspiracy theories. The content of rumors may vary in terms of the language used, the presence of multimedia elements (such as images or videos), and the level of detail provided.
Rumors often follow specific narrative structures or storytelling patterns. They may contain elements such as dramatic twists, personal anecdotes, vivid imagery, or appeals to emotion. Analyzing the narrative structure of rumors can provide insights into their persuasion power and potential impact on audiences \cite{Shelke2022}.

\smallskip
\subsubsection{Limited ground truth}
Obtaining labeled datasets for training and evaluating rumor detection algorithms can be difficult owing to the subjective nature of labeling rumors. Ground truth data may be sparse or unreliable, hindering the development of accurate detection models.
Ground truth refers to labeled data that accurately indicates whether a piece of information is a rumor or not. This labeled data is crucial for training and evaluating rumor-detection models \cite{Zafarani15}.
Obtaining reliable ground truth data for rumor detection is challenging due to the subjective nature of labeling rumors. Different individuals may interpret the same piece of information differently, leading to inconsistencies in labeling. 
Limited ground truth data can hinder the development and evaluation of rumor-detection models. Without a sufficient amount of labeled data, it is challenging to train accurate and robust detection algorithms. Furthermore, biased or unreliable ground-truth labels may lead to the development of models that perform poorly in real-world scenarios.

\smallskip
\subsubsection{Temporal dynamics}
Rumors evolve over time, with information addition, mutation, and sometimes, debunking or correction \cite{Gelardi2021}. Tracking the temporal dynamics of rumor propagation requires techniques capable of capturing changes in content, context, and user interactions over time. Temporal dynamics refer to the changes that occur in rumors over time as they propagate through social networks. These changes may include modifications to the content of the rumor, shifts in sentiment or tone, and variations in the propagation speed and reach.
 Rumors evolve rapidly, with new information being added, modified, or debunked as they spread. This temporal evolution makes it challenging to track and detect rumors accurately \cite{Kwon2017}. Detection models must be capable of adapting to these changes and updating their assessments in real time.
Failing to account for temporal dynamics can lead to inaccurate or outdated detection results. Rumors that were true at one point in time may become false as new evidence emerges, while rumors that were false initially may gain credibility over time. Detection models must be able to capture these temporal nuances to provide reliable assessments of rumor veracity.

\smallskip
\subsubsection{Platform-specific challenges}

Platform-specific challenges in rumor detection refer to the unique obstacles and complexities associated with detecting rumors on different social media platforms \cite{Chang2024}.
Each social media platform has its own application programming interface (API) with specific limitations on data access, including rate limits, data format restrictions, and access to historical data.
Different social media platforms support various types of content formats, including text, images, videos, links, and hashtags. Rumors may manifest differently across platforms, requiring flexible detection techniques capable of handling diverse content formats.
Rumor-detection models must be adaptable to different content types and formats to identify rumors across multiple platforms effectively. Failure to account for content variation may result in missed rumors or inaccurate detection results.
Social media platforms have distinct network structures, user demographics, and engagement metrics that influence the spread of rumors. Some platforms may emphasize follower-based networks, while others rely on friend-based connections or interest-based communities.

\begin{table*}[t!]
\renewcommand{\arraystretch}{1.3}
\caption{Summary of Research Challenges and Possible Directions.} \label{tab:direction} \centering \begin{tabular}{p{2cm}|p{6.5cm}|p{7.5cm}}
\hline      & ~~~~~~~~~~~~~~~~~~~~~\textbf{Issues \& Challenges} & ~~~~~~~~~~~~~~~~~~~~ \textbf{Possible Directions} \\
\hline \textbf{Rumor Detection} & Data volume and velocity & Design a method to first apply a technique to filter information noise from large amounts of data and then extract unique features that only rumors have. Utilize sensors at important nodes in the network to enable early detection of rapidly spreading rumors\\
\cline{2-2} \cline{3-2} & Variability in rumor characteristics & Among the various features of rumors, extract the features most heavily used in rumor detection and make a decision based on this. In other words, assign weights to features and apply them to rumor detection\\
\cline{2-2} \cline{3-2} & Limited ground truth & Simultaneously learn labeled data and unlabeled data, then derive a label method for unlabeled data and design a detection algorithm based on this. For example, in machine learning, self-supervised learning techniques are applied to derive interpretation of unlabeled data\\
\cline{2-2} \cline{3-2} & Temporal dynamics & Consider an on-line algorithm that can analyze the context in real time to determine whether it is a rumor or not. To achieve this, set an appropriate time-period and perform detection tailored to each situation at each time. Then consider how to estimate the actual rumor based on majority vote or other estimation methods.\\
\cline{2-2} \cline{3-2} & Platform-specific challenges & Develop an algorithm that can analyze the characteristics of different social platforms, derive detection methods for them, and then fuse them\\
\hline  \textbf{Rumor Source Detection}  & Anonymity and pseudonymity & First, estimate the source node based on the structure of the infection graph or infection path of the information over the network and then proceed with the task of solving anonymity using various network anonymity detection algorithms \\
\cline{2-2} \cline{3-2} & Network dynamics & Design a detection algorithm based on the assumption that the network is static for a certain period of time. Then, design a method to track the initial spreader for the entire time based on the source node estimated for each time period. Or, consider a technique to trace back to the source based on message routing information in a dynamic network\\
\cline{2-2} \cline{3-2} & Legal and ethical considerations & If the information is determined to be a rumor, estimate the source based on the structure of the infection graph. Otherwise, consider a technique to hide the source for privacy\\
\cline{2-2} \cline{3-2} & Cross-platform propagation & First analyze the characteristics of information spread by platform and parameterize this to estimate the source based on the rumor spread pattern in the case of multi-platforms. For example, algorithm design that takes into account changes in the propagation rate on a specific platform or changes in the pattern of spreading rumors\\
 \hline
 \textbf{Joint Detection}  & Integration of multi-modal data & To detect rumors, apply an algorithm that can effectively utilize multimodal data and perform source detection by estimating the propagation path based on this\\
 \cline{2-2} \cline{3-2} & Temporal misalignment & Analyze the temporal dynamics of rumors and the dynamics of network structures and propose a method to detect both together based on the characteristics of multiple platforms\\
 \cline{2-2} \cline{3-2} & User behavior Modeling and adversarial manipulation & Parameterize the credibility and reliability of the user, analyze user behavior based on this, proceed with rumor detection, and estimate the source based on the propagation pattern. For example, rumor detection is carried out simultaneously based on the assumption that information spread from sources with low reliability is likely to be a rumor\\
 \cline{2-2} \cline{3-2} & Scalability/Efficiency/Interpretability & By analyzing the characteristics of rumor source nodes well in advance, devise an efficient approach that can interactively solve the problems of rumor detection and source detection\\
  \hline
 \end{tabular} \end{table*}

Addressing these challenges requires interdisciplinary approaches that combine techniques from NLP, machine learning, network science, sociology, and psychology. Developing robust rumor-detection systems requires ongoing research, collaboration, and adaptation to evolving misinformation tactics and social media dynamics.

\subsection{Challenges Associated with Rumor-Source Detection}

Detecting the source of rumors presents its own set of challenges, distinct from those encountered in rumor detection. Here are some of the key challenges for rumor-source detection:

\smallskip
\subsubsection{Anonymity and pseudonymity}
Social media platforms often allow users to create accounts anonymously or under pseudonyms, making it difficult to trace the true identity of individuals spreading rumors. Determining the actual source of misinformation when users operate under aliases can be challenging. Anonymity and pseudonymity present significant challenges in rumor-source detection, as they allow individuals to conceal their true identities while disseminating information online \cite{Peddinti2014}. 
Anonymity refers to the condition of being unknown or unidentified. On social media platforms, users may choose to remain anonymous by not disclosing their real names or personal information.
Anonymity makes it difficult to trace the source of rumors back to specific individuals or entities. Without identifiable information such as usernames, profile pictures, or account metadata, it becomes challenging to attribute rumors to their originators.
Rumors spread by anonymous users may be more difficult to debunk or refute, as there is no clear source to hold accountable for spreading false information. Additionally, anonymous individuals may feel emboldened to spread rumors without fear of consequences, contributing to the proliferation of misinformation online.

\smallskip
\subsubsection{Network dynamics}
Rumors propagate through complex networks of social interactions, with information flowing through multiple pathways and nodes. Analyzing the topology of social networks, identifying influential users, and tracing the diffusion patterns of rumors require sophisticated network-analysis techniques capable of handling large-scale, dynamic data \cite{Jiang19}.
Network dynamics and complexity play crucial roles in rumor-source detection, as rumors propagate through complex social networks shaped by diverse interactions and relationships among users. 
Social networks consist of nodes which representing individuals or entities and edges that representing connections or interactions between nodes. The structure of the network, including its size, density, centrality, and modularity, influences the spread of rumors.
Detecting rumor sources requires understanding the structural properties of social networks and identifying the key nodes involved in rumor propagation. However, networks may exhibit heterogeneity, with some nodes possessing higher degrees of influence or connectivity than others.
Rumor-source detection algorithms must account for the network structure to prioritize influential nodes and trace the paths of rumor propagation \cite{Choi20d}. Failure to consider the network dynamics may result in inaccurate attribution of rumors or ineffective strategies for controlling their spread.
Social networks evolve over time, with connections forming, breaking, and evolving dynamically. Rumors also evolve temporally, with information being added, modified, or debunked as they spread.
Rumor-source detection algorithms must capture the temporal dynamics of social networks and rumors to track changes in network structure, content, and engagement over time.
Analyzing temporal dynamics can provide insights into the emergence, evolution, and persistence of rumors within social networks. Detection algorithms must adapt to changes in network topology and content characteristics to maintain accurate attribution of rumors and sources over time.

\smallskip
\subsubsection{Legal and ethical considerations}

Investigating the source of rumors may raise legal and ethical concerns related to user privacy, data protection, and freedom of expression. Balancing the need to combat misinformation with respect for users' rights and freedom poses challenges for rumor-source detection efforts.
Legal and ethical considerations are paramount in rumor-source detection \cite{Isabelle2021}, as they involve identifying individuals or entities responsible for spreading misinformation online. 
Rumor-source detection often involves collecting and analyzing data from social media platforms, which may raise privacy and data protection concerns. Laws such as the General Data Protection Regulation (GDPR) in the European Union and the California Consumer Privacy Act (CCPA) in the United States impose strict requirements on collecting, processing, and storing personal data.
 Respecting user privacy and data protection rights is essential in rumor-source detection. Researchers and practitioners must ensure compliance with the relevant privacy regulations and obtain informed consent from users when collecting and analyzing data.

\smallskip
\subsubsection{Cross-platform propagation}
Rumors often spread across multiple social media platforms, blogs, forums, and online communities \cite{Yaming2019}. Tracking the cross-platform propagation of rumors and coordinating detection efforts across different platforms presents technical, logistical, and jurisdictional challenges.
Cross-platform propagation poses several challenges for rumor-source detection:
Tracking rumors across multiple platforms requires integrating data from disparate sources, each with its own data format, access restrictions, and API limitations.
Rumors may manifest differently on different platforms, with content format, language, and user engagement variations. Detecting rumor sources across platforms requires algorithms capable of handling contextual variability.
Coordinating detection efforts across platforms and collaborating with platform providers, fact-checkers, and researchers is essential for comprehensive and effective rumor-source detection.
Rumors may spread across platforms hosted in different jurisdictions, each with its own legal and regulatory framework. Addressing jurisdictional issues requires international cooperation and adherence to applicable laws and regulations.

\subsection{Challenges for Joint Detection}
Detecting both rumors and their sources jointly presents several interconnected challenges, as it involves analyzing the content, propagation patterns, and network dynamics of misinformation. Here are some of the key challenges for rumor detection and source detection when addressed jointly:

\smallskip
\subsubsection{Integration of multimodal data}
Jointly detecting rumors and their sources requires integrating information from multiple sources, including textual content, user interactions, network structure, and temporal dynamics \cite{Yang2023}. Effectively combining diverse data modalities and extracting meaningful features from heterogeneous sources pose challenges for developing integrated detection models.
Rumor detection relies on analyzing the semantic content and contextual cues of messages, while source detection involves tracing the origin and propagation paths of rumors through social networks. Understanding the semantic and contextual relationships between rumors and their sources requires advanced NLP techniques, sentiment analysis, and context-aware modeling.

\smallskip
\subsubsection{Temporal misalignment}
As described before, rumors may evolve over time, with information addition, mutation, and being debunked or corrected. Tracing the temporal dynamics of rumor propagation and source attribution requires aligning timestamps, tracking changes in content and context, and accounting for delays and latency in information dissemination \cite{Wang2021}.
Rumors often spread across multiple social media platforms, blogs, forums, and online communities. Jointly detecting rumors and their sources across different platforms requires cross-platform data integration, interoperability, and coordination to ensure comprehensive coverage and accurate attribution.

\smallskip
\subsubsection{User behavior modeling and adversarial manipulation}
Understanding the motivations, intentions, and behaviors of the users spreading rumors is crucial for both rumor detection and source detection \cite{Chen2018}. Modeling user-engagement patterns, social influence dynamics, and information-sharing behaviors across networks poses challenges for capturing the complex interplay between individual actions and collective phenomena.
Malicious actors may exploit vulnerabilities in detection algorithms, manipulate content, or engage in coordinated campaigns to evade detection and conceal their identity as rumor sources. Developing robust detection mechanisms resilient to adversarial manipulation requires adversarial training, outlier detection, and anomaly detection techniques.

\smallskip
\subsubsection{Scalability/Efficiency/Interpretability}
Detecting rumors and their sources in real-time across large-scale social media platforms requires scalable, efficient, and resource-effective detection algorithms. Addressing scalability challenges involves optimizing computational performance, parallelizing processing tasks, and leveraging distributed computing infrastructure.
Providing interpretable and transparent explanations for detection decisions is essential for building trust and accountability in detection systems \cite{Lyu2022}. Balancing the need for model transparency with the complexity of joint-detection algorithms poses challenges for ensuring the interpretability and explainability of detection results.

\section{Conclusion}
\label{sec:conclusion}
The paper provided a rigorous and in-depth survey on the topics of
rumor and rumor-source detection on social networks, along with a joint consideration of these problems. The paper
first introduced each of the three problems with formal problem formulations 
and described the algorithms proposed to solve the problems.
We also introduced the problem of hiding rumors and rumor sources and summarized the related papers. In addition, overall, the limitations arising from the rumor-detection, source-detection, and joint problems were explained, and opinions on future directions were presented.

\balance
{
\renewcommand{\baselinestretch}{0.93}
\bibliographystyle{IEEEtran}

}


\end{document}